\documentclass{elsart5p}

\usepackage{graphicx}
\usepackage{amssymb}

\begin{document}

\begin{frontmatter}



\title{Certified Rational Parametric Approximation of Real Algebraic Space Curves with Local Generic Position Method}

%\footnotetext{The work is partially supported by NKBRPC
%(2011CB302400), NSFC Grants (60821002, 11001258), and China-France
%cooperation project EXACTA (60911130369).}

\author{Jin-San Cheng, Kai Jin and Xiao-Shan Gao}

\address{KLMM, Institute of Systems Science, AMSS, CAS, Beijing 100190, China}

\author{Daniel Lazard}

\address{Universite Pierre et Marie Curie, INRIA Paris-Rocquencourt Research Center}
\address{jcheng@amss.ac.cn,xgao@mmrc.iss.ac.cn,jinkaijl@163.com,daniel.lazard@lip6.fr}

\begin{abstract}
In this paper, an algorithm to compute a certified $G^1$ rational
parametric approximation for algebraic space curves is given by
extending the local generic position method for solving zero
dimensional polynomial equation systems to the case of dimension
one. By certified, we mean the approximation curve and the original
curve have the same topology and their Hausdauff distance is smaller
than a given precision. Thus, the method also gives a new algorithm
to compute the topology for space algebraic curves.
%
%The piecewise approximation curve segments have one of the following
%forms: $(x,
%a\,x+b+\frac{c}{d\,x+1},a_1\,x+b_1+\frac{c_1}{d_1\,x+1}+\frac{c_2}{d_2\,x+1})$,
%$(a\,y+b+\frac{c}{d\,y+1},y,a_1\,y+b_1+\frac{c_1}{d_1\,y+1}+\frac{c_2}{d_2\,y+1})$.
%
The main advantage of the algorithm, inhering from the local generic
method, is that topology computation and approximation for a space
curve is directly reduced to the same tasks for two plane curves. In
particular, the error bound of the approximation space curve is
obtained from the error bounds of the approximation plane curves
explicitly. Nontrivial examples are used to show the effectivity of
the method.\end{abstract}

\begin{keyword}
Real algebraic space curve\sep topology preserving \sep rational approximation parametrization \sep local generic position
\end{keyword}
\end{frontmatter}

\newcommand{\newOld}[2]{#1}

\renewcommand\refname{References and Notes}
\renewcommand{\abstractname}{Abstract}
\renewcommand{\thefootnote}{\fnsymbol{footnote}}

\newtheorem{theorem}{Theorem}
\newtheorem{corollary}[theorem]{Corollary}
\newtheorem{algo}[theorem]{Algorithm}
\newtheorem{lemma}[theorem]{Lemma}

\def\R{{\mathbb{R}}}
\def\Z{{\mathbb{Z}}}
\def\C{{\mathbb{C}}}
\def\Cc{{\mathcal{C}}}
\def\Ss{{\mathcal{S}}}
\def\Q{{\mathbb{Q}}}
\def\I{{\mathbb{I}}}
\def\B{{\bf B}}
\def\A{{\mathcal{A}}}
\def\F{{\mathcal{F}}}
\def\Bc{{\mathcal{B}}}
\def\S{{\mathbb{S}}}
\def\i{{\mathfrak{i}}}

\def\PS{ {\mathcal{P}} }
\def\GB{{\mathcal{G}}}
\def\IS{{\mathcal{I}}}
\def\IS{{\mathcal{L}}}
\def\BS{{\mathbf{BS}}}

\def\CB{{\C_{\B}}}

\def\RB{\hbox{\rm{RB}}}
\def\Dis{\hbox{\rm{Dis}}}

\def\Re{\hbox{\rm{Re}}}
\def\Im{\hbox{\rm{Im}}}

\def\bref#1{(\ref{#1})}
% main text
\section{Introduction}
\label{intro}

Algebraic space curves have many applications in computer aided geometric design, computer aided design, and geometric modeling. For example, the algebraic space curves defined by two quadrics are widely used in geometric modeling. One can have an exact parametrization for these algebraic space curves. However, exact parametrization representation for general algebraic space curves do not exist. And usually, we are interested in a rational parametric representation. So the use of approximate techniques is unavoidable for parametrization of algebraic space curves. Some approximate techniques are able to reproduce exact rational parameterizations, if those are available. Otherwise, one usually approximates an algebraic space curves with piecewise rational curves under a given precision. Moreover, sometimes, one requires that the approximation curves preserve the topology of the original algebraic space curve. We call the approximation {\bf certified} (at precision $\epsilon$) if it has the same topology as the original one and the Hausdorff distance between the curve and its approximation is upper-bounded (by $\epsilon$) simultaneously.

There are several difficulties for approximate parametrization of algebraic space curves. The first one is to preserve the topology of the algebraic space curve. In fact, there already exist some related work of computing the topology of algebraic space curves, reduced or non-reduced, for example \cite{as,dao,el,gat,owen}. Most of them require the curve to be in a generic position. For the space curves which are not in a generic position, one need to take a coordinate transformation such that the new space curve is in a generic position. Thus some geometric information of the original space curve is lost. Some non-singular critical points of the new space curve may not correspond to the non-singular critical points of the original space curve. One needs additional computation to get these points in the original coordinate system. Subdivision method can preserve the topology of the curve in a
theoretical sense. But it is rather difficult to reach the required bound in practice currently \cite{yap, liang}. Even if one gets the topology of the given curve, the approximation curve may have different topology as the original curve when two or more curve segments are very close( see Figure 2).
The second difficulty is the error control of the approximation curve. Some error functions is reliable but it is not easy to compute in practice, for example \cite{chuang}. We need to find a reliable and efficient method to control the error during the approximation.
The third one is the continuity of the approximation curve. We usually require the approximation to be $C^1$(or $G^1$)-continuous or higher in practice. Doing so, we need compute the tangent directions of algebraic space curve at some points. It is not a trivial task especially when the component considered is {\it non-reduced}. Its tangent direction can not be decided by the normal directions $(\frac{\partial f}{\partial x},\frac{\partial f}{\partial y},\frac{\partial f}{\partial z})$, $(\frac{\partial g}{\partial x},\frac{\partial g}{\partial y},\frac{\partial g}{\partial z})$ of the two surfaces $f=0$ and $g=0$ at the given point. In the non-reduced case, two normal directions are parallel or at least one of them does not exist at the point.

There exist nice work about approximating of algebraic space curves. Exact parametrization of the intersection of algebraic surfaces is obtained in \cite{abh,biz,nancy1,nancy2,nancy3,gaochou,van,sch,sendra,tu,wang}. Of course, it is topology preserving. The approximation of the intersection of generic algebraic surfaces with numeric method is also considered \cite{baj1,baj,kri,pat,pra}. Usually, numeric method cannot guarantee the topology of the original algebraic space curve.
%A property of numeric method is that the topology of the original space curve is not guaranteed.
% In \cite{skyu}, the author pointed out the challenges of intersection problem, especially for the intersection of parametric surfaces.

In \cite{bela}, the authors considered approximating of the regular algebraic space curves with circular arcs by numeric method combining with the subdivision method. It works well for low degree algebraic space curves. In \cite{gaoli}, the authors present an algorithm to approximate an irreducible space curves under a given precision. It based on the fact that there exists a birational map between the projection curve $\Cc$ for some direction and the irreducible algebraic space curve $S$. % in some direction to $S$.
But an irreducible decomposition of a given two polynomials system is not an easy task. And we need to consider the intersection of two or more irreducible space curves after we decompose a reducible space curve. Other type of intersection problem of surfaces can be found in \cite{pat,pra} and related references.

In \cite{biz}, the authors present an algorithm to approximate an algebraic space curve, defined by $f=g=0$, in the generic position with Ferguson's cubic $\mathbf{p}(t)=(x(t),y(t),z(t)), t\in[0,1]$ and by minimizing an integral to control the error.
%$$\int_0^1(\frac{f^2(x(t),y(t),z(t))}{\|\nabla f(x(t),y(t),z(t))\|^2}+\frac{g^2(x(t),y(t),z(t))}{\|\nabla g(x(t),y(t),z(t))\|^2})dt.$$
They compute the topology of the space curve at first, so it is topology preserving. But they do not check whether the approximation curve exactly preserves the topology of the original space curve. And it works well for regular space curve. From the formula above, we can find that if some segments of the algebraic space curve is not regular, the method may fail.
 %since $\nabla f(x(t),y(t),z(t))=0$ or $\nabla g(x(t),y(t),z(t))=0$ may happen.

In \cite{rue}, the authors consider the irreducible algebraic space curve in generic position such that its projection is birational. They use a genus 0 plane algebraic curve to approximate the projection plane curve under a given precision if it exists. Thus they have a rational approximation space curve for the original space curve. The method is not topology preserving.

In this paper, we present a new algorithm to compute a certified
$G^1$ rational parametric approximation for algebraic space curves,
which solves the three difficulties mentioned above nicely. The
algorithm is certified in the sense that the approximation curve and
the original curve have the same topology and their Hausdauff
distance is smaller than a given precision. The algorithm works for
algebraic space curves which need not to be regular or in generic
positions. The key idea is to extend the local generic position
method \cite{lgp2,lgp} for zero-dimensional polynomial systems to
one-dimensional algebraic space curves. The algorithm consists of
four major steps.

Firstly, the space curve $\Ss$, which is the intersection of
$f(x,y,z)=0$ and $g(x,y,z)=0$, is projected to the $xy$-plane as a
plane curve $\Cc_1$ and $\Cc_1$ is approximated piecewisely with
functions of the form $h_1(x),x\in[a,b]$.

Secondly,  we find a number $s>0$ such that under the shear
transformation $\varphi: (x,y,z)\rightarrow (x,y+s\,z,z)$,
$\varphi(f)$ and $\varphi(g)$ are in a generic position in the sense
that there is a one to one correspondence between the curve segments
of $\Ss$ and that of their projection curve $\Cc_2$ to the
$xy$-plane.
The plane curve $\Cc_2$ is also approximated piecewisely with
functions of the form $h_2(x),x\in[a,b]$.

Thirdly, we choose $s$ such that $\Cc_2$ is in a {\bf local
generic position} to $\Cc_1$ in the following sense.
\begin{itemize}
\item The plane curves $\Cc_1$ and $\Cc_2$ can be divided into
segments such that each segment of $\Cc_2$ corresponds to a segment
of $\Cc_1$.

\item Let $h_1(x), h_2(x),x\in[a,b]$ be the approximations for a
segment $C_1$ of $\Cc_1$ and the corresponding curve segment $C_2$ of
$\Cc_2$ with precisions $\epsilon_1$ and $\epsilon_2$ respectively.
Then the space curve segment $S$ corresponding to $C_2$ can be
approximated by $(x,h_1(x),\frac{h_2(x)-h_1(x)}{s})$ with precision
$\sqrt{s^2\epsilon_1^2+(\epsilon_1+\epsilon_2)^2}/s$.
\end{itemize}

In other words, if $\Cc_2$ is in a local generic position to
$\Cc_1$, then each segment of the space curve can be represented as
a linear combination of corresponding segments of $\Cc_1$ and
$\Cc_2$. As a consequence, a certified parametrization for
the space curve can be computed from that of $\Cc_1$ and $\Cc_2$
directly. This step is the main contribution of the paper.

Finally, we show that a plane curve can be approximated such that
the piecewise approximation curve for the space curve has $G^1$
continuity and usually has the following forms: $(x,
a\,x+b+\frac{c}{d\,x+1},a_1\,x+b_1+\frac{c_1}{d_1\,x+1}+\frac{c_2}{d_2\,x+1})$,
$(a\,y+b+\frac{c}{d\,y+1},y,a_1\,y+b_1+\frac{c_1}{d_1\,y+1}+\frac{c_2}{d_2\,y+1})$.

The topology of the space curve is obtained directly from the two
projection steps. Thus this new method can not only compute the
topology of the space curve but approximate the algebraic space
curve under any given precision.

The paper is organized as below. In the next section, we will
consider the certified approximation of plane algebraic curve under
a given precision. In Section 3, we will show the theory and
algorithm for certified approximation of algebraic space curves. In
Section 4, we will show some examples to illustrate the effectivity
of our method. We draw a conclusion in the last section.

\section{Approximate parametrization of plane algebraic curves }
%In this section, we will solve the following problem: \\
Given a plane algebraic curve defined by a square free polynomial $f\in \Q[x,y]$, our aim is to give a piecewise $C^1$-continuous approximation of $\Cc=\{(x,y)\in \R^2|f(x,y)=0\}$ in a given box $B=\{(x,y)\in \R^2|a\le x\le b, c\le y\le d\}$ such that each piece of the approximation curve has the form $(x,h(x))$ and the approximation error is bounded by a given precision $\epsilon>0$, where $\Q,\R$ are the fields of rational numbers and real numbers, respectively. And the whole approximation curve has the same topology as $\Cc$.

%In \cite{gaoli}, an algorithm is proposed to give a global approximation of a real plane algebraic curve with rational
%quadratic B-spline curves. The main steps in the method are topology determination, curve segmentation,
%segment approximation and curve tracing. We will recall the methods of approximate parametrization of plane algebraic curve and modify some existing steps for our own purposes by introducing some new techniques.
%

\subsection{Notations}
In this subsection, we will introduce some notations. %Some of the notations are similar as the ones in \cite{gaoli}.

A point $P = (x_0, y_0)$ is said to be a {\bf singular point} on $\Cc: f(x,y)=0$ if $f(x_0, y_0) = f_x(x_0, y_0) = f_y(x_0, y_0) = 0$, where $f_x=\frac{\partial f}{\partial x}, f_y=\frac{\partial f}{\partial y}$. A non-singular point is called a {\bf regular point}.
An {\bf $x$-critical} (A {\bf $y$-critical}) point $P = (x_0, y_0)$ is a point satisfying $f(x_0, y_0) = f_y(x_0, y_0) = 0 $ ($f(x_0, y_0) = f_x(x_0, y_0) = 0$). So a singular point is both $x$-critical and $y$-critical points.
The {\bf inflexion points} or {\bf flexes} of $\Cc$ are its non-singular points satisfying its Hession equation $H(f) = 0$ (see \cite{walker}).

A {\bf regular curve segment} $C$ of $\Cc$ is a connected part of $\Cc$ with two endpoints $P_0(x_0,y_0)$ and $P_1(x_1,y_1)$ ($x_0\neq x_1$, both $P_0,P_1$ are bounded) and there are no $x$-critical points, $y$-critical points, flex on $C$ except for $P_0,P_1$. So a regular curve segment is convex, monotonous w.r.t. $x$ or $y$, and inside a triangle defined by its endpoints and their tangent directions.
Let $\Delta$ be the triangle defined by $P_0,P_1$ and their tangent lines.
An endpoint of a regular curve segment is called a {\bf vertical tangent point}, {\bf VT} point for short, if the regular curve segment has a vertical tangent line at this endpoint.
%
%The {\bf left (right) endpoint} is the one with smaller (larger) $x$-coordinate.

%A parametric curve  $(x_1(t),\ldots,x_n(t)), t\in [a,b]$ is said to be {\bf $C^1$-continuous} if for any $t_0\in (a,b)$, we have
%$$\lim_{t\rightarrow t_0^+}(x_i(t))=\lim_{t\rightarrow t_0^-}(x_i(t))=x_i(t_0), \lim_{t\rightarrow t_0^+}(\frac{\partial x_i(t)}{\partial t})=\lim_{t\rightarrow t_0^-}(\frac{\partial x_i(t)}{\partial t})=\frac{\partial x_i(t)}{\partial t}|_{t=t_0}, i=1,\ldots,n.$$

A parametric curve is said to be {\bf $C^1$-continuous} ({\bf $G^1$-continuous}) if the curves are joined and the first derivatives are continuous (the curves also share a common tangent direction at the join point).

\subsection{Curve segmentation of a real plane algebraic curve}
In this subsection, we will show how to divide a plane curve inside a box $B$, denoted as $\Cc_B$, into regular curve segments with the form $[P_0(x_0,y_0), P_1(x_1,y_1), T_0(1,k_0), T_1(1,k1)]$, where $P_0,P_1$ are endpoints and $T_0,T_1$ are tangent directions at the endpoints.

We will follow the steps below.
%\begin{enumerate}
%\item Compute the topology of $\Cc_B$.
%\item Compute all the flexes, $x$-critical and $y$-critical points of $\Cc_B$.
%\item Split $\Cc_B$ into regular curve segments.
%\item Compute the tangent direction of the endpoints of the regular curve segments.
%\end{enumerate}

At first, Compute the topology of $\Cc_B$. There are many related work to solve this problem, such as \cite{alb,arn,berb,cheng,eig,hong,sak,sei}. Some methods work well, but they need a coordinate system transformation. We prefer the methods which do not take a coordinate system transformation.

Second, Compute all the flexes, $x$-critical and $y$-critical points of $\Cc_B$. $y$-critical points are computed before. $x$-critical points and flexes can be obtained by solving the corresponding equations.

Third, we split the plane curve into regular curve segments at $x$($y$)-critical points or flexes of $\Cc_B$. An easy way to solve the problem is as follows. For all the $x$-coordinates of $x$-critical points and flexes, lifting them to split $\Cc_B$  at these intersection points. We can find all these endpoints.

Finally, we represent the tangent direction of any non-VT point as $(1,k), k\in \R\setminus\{+\infty,-\infty\}$. It is convenient for our approximation. The tangent direction of a VT point is defined to be $(1,\infty)$.
%Though we can compute the tangent directions of singular points following the algorithm in \cite{gaoli}. We prefer a direct way to get the tangent directions of singularities by approximating as below.\\
\\
\noindent{\bf Tangent direction computation of singularities.}
We compute the tangent direction of a point close to a singularity on the regular curve segment to replace the tangent direction of the singularity. It is easy to compute. Let $P(\alpha,\beta)$ be a singularity of a planar algebraic curve $h(x,y)=0$ and $C: (x, \tilde{y}(x)), x\in[\alpha,\gamma]$ a regular curve segment originating from right side (left side is similar) of $P$. Then the tangent direction of $C$ at $P$ is $(1,t)=(1,\lim_{x\rightarrow \alpha^+}\tilde{y}(x)).$
 In practice, we can take some point very close to $P$ on $C$, which is a regular point. Let $[a,b]$ be the isolating interval of $\alpha(\neq a,\neq b)$. We can use the tangent direction of some regular point to replace the tangent direction of $C$ at $P$. For instance, $(1,\frac{\partial \tilde{y}(b)}{\partial x})=(1,\frac{f_x}{f_y})$ can be regarded as the tangent direction of $C$ at $P$.
 For the regular curve segments shall the same tangent direction, we can take the average value of them as their tangent directions, and the center of the isolating box of $P$ as the singularity, as shown in Figure 1. %\ref{fig-tangentapp}.
 When $C$ has a vertical tangent direction, $t=\infty$. If $|\frac{\partial \tilde{y}(b)}{\partial x}|>N$, for example, $N=100$, we can regard the regular curve segment have a vertical tangent direction.

If we cannot distinguish the tangent directions of two groups of regular curve segments, we can refine $[a,b]$ to a narrower one and recompute the tangent directions again until we can distinguish them or they are less than some given bounded value $\tau$ such that $|k-k'|<\tau$, where $k,k'$ are tangent directions.

%In this subsection, we will decompose a real algebraic curve inside a box into regular curve segments such that the singularities, flexes, $x$-critical points, $y$-critical points are endpoints of some regular curve segments, and the graph formed by these regular curve segments has the same topology as the real algebraic curve inside the box. We also compute the tangent line of these regular curve segments. The tangent vector at one endpoint is $(1,k)$, where $k\in[-\infty,+\infty]$. It means that the regular curve segment has a vertical tangent line at the endpoint when $k=-\infty$ or $k=+\infty$.
%

    \begin{figure}[hb]
\begin{minipage}{0.234\textwidth}\label{fig-tangentapp}
\centering
\includegraphics[height=25mm]{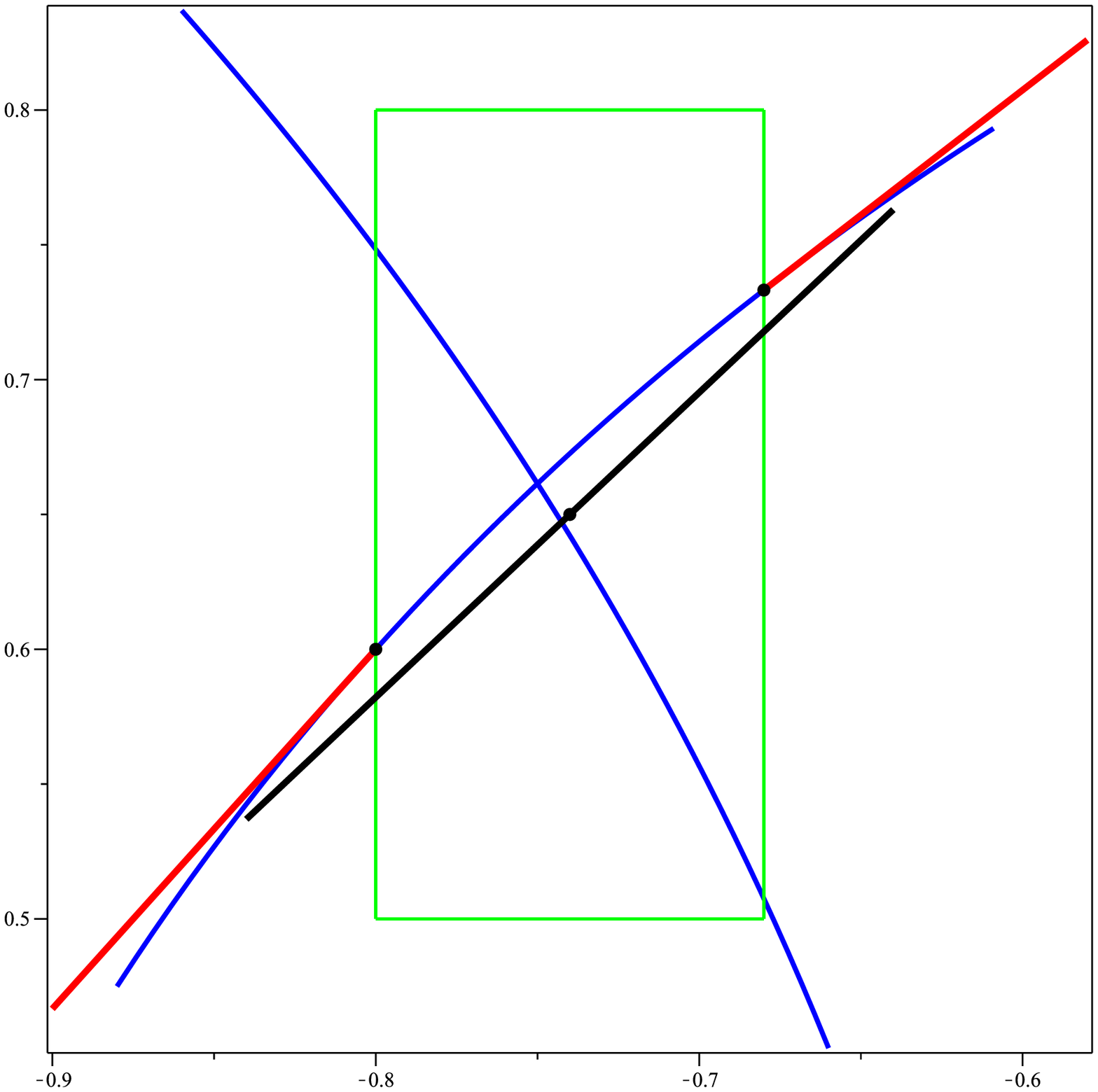}\caption{Approximate the tangent direction of regular curve segments at a singularity}
\end{minipage}
\begin{minipage}{0.234\textwidth}\label{fig-apptoperror}
\centering
\includegraphics[height=22mm]{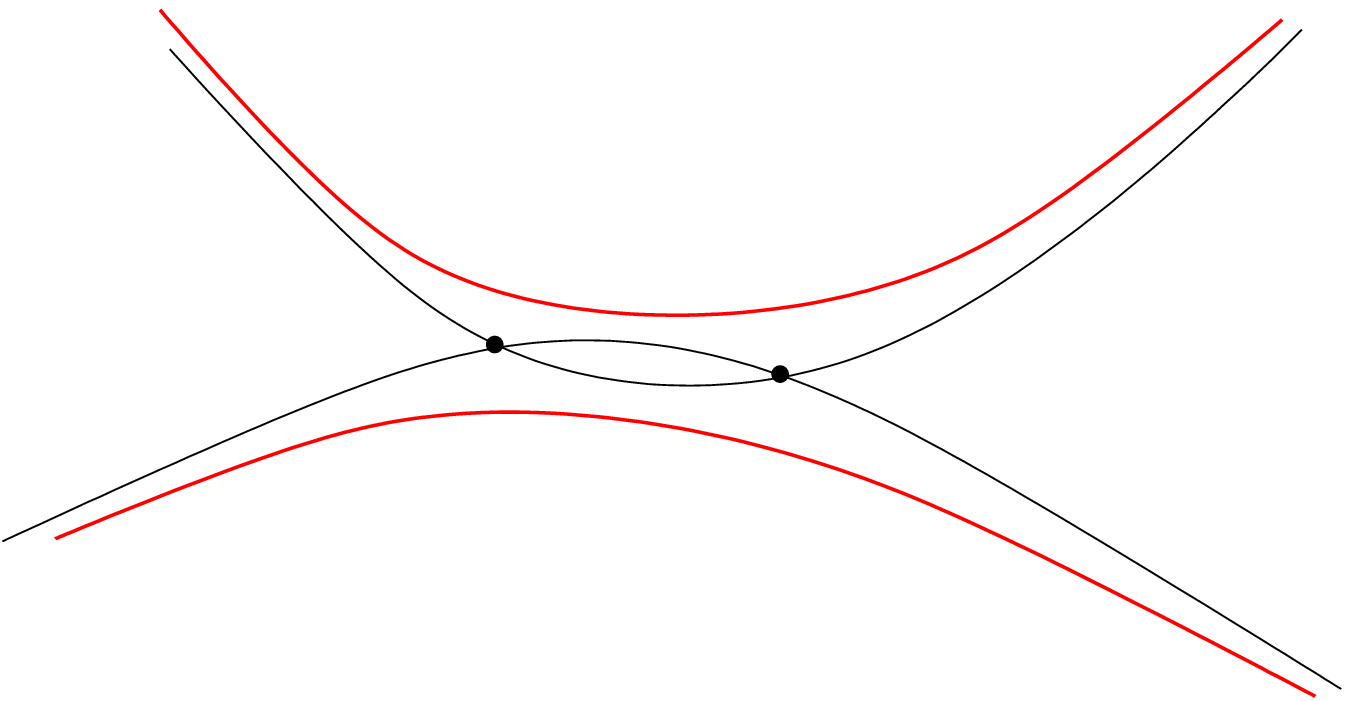}\caption{The approximation curves change the topology of original ones}
\end{minipage}
    \end{figure}

\subsection{Approximation of a regular curve segment}

We will give an approximate parametrization of real plane algebraic curves. Though Gao and Li \cite{gaoli} have obtained a rational quadratic approximation of real plane algebraic curves with B-splines, we need to derive a piecewise approximation curve as $(x, h(x))$ of real plane algebraic curves in order to approximately parameterize real space algebraic curves in a different way.  Let $C$ be the regular curve segment defined by two points $P_0(x_0,y_0),P_1(x_1,y_1)$.  We divide the approximation problem into two cases.
%In the rest of this paper, we require the regular curve segment has one critical point ($x$ or $y$-critical point), otherwise we can divide the regular curve segment into two regular curve segments.

\noindent{\bf $C$ not containing a VT point.}
Given a regular curve segment $C$, its two endpoints $P_0(x_0,y_0),P_1(x_1,y_1)$ does not contain a VT point. And the tangent directions of the regular curve segment at $P_0, P_1$ are $(1,k_0), (1,k_1)$, respectively. We will construct an explicit rational quadratic function $Y_1(x)$ to approximate $C$ such that $Y_1(x_i)=y_i, Y_1'(x_i)=k_i, i=0,1$, where $Y_1'(x)=\frac{\partial Y_1}{\partial x}(x)$.

Assuming that \begin{equation}\label{eq-y1} Y_1(x)=\frac{a\, x^2+b\,x+c}{d\,x+1}\end{equation} and $x_0=0, x_1=1$, we have
\begin{eqnarray*}
Y_1(0)=y_0, Y_1(1)=y_1,
Y_1'(0)=k_0, Y_1'(1)=k_1.
\end{eqnarray*}

Solving $a,b,c,d$ from the equations above, we have
\begin{eqnarray*}
a&=&{\frac {-2\,{\it y_0}\,{\it y_1}+{{\it y_0}}^{2}-{\it k_0}\,{\it k_1}
+{{\it y_1}}^{2}}{-{\it y_1}+{\it k_1}+{\it y_0}}},\\
b&=&-{\frac {-2\,{\it y_0
}\,{\it y_1}+{\it y_0}\,{\it k_1}+2\,{{\it y_0}}^{2}+{\it k_0}\,{\it y_1}-
{\it k_0}\,{\it k_1}}{-{\it y_1}+{\it k_1}+{\it y_0}}},\\
c&=&{\it y_0},\\
d&=&-{
\frac {-2\,{\it y_1}+{\it k_0}+{\it k_1}+2\,{\it y_0}}{-{\it y_1}+{\it
k_1}+{\it y_0}}}.
\end{eqnarray*}

%{\scriptsize
%\begin{eqnarray*}
%{\it b_1}&=&2\,{{\it y_1}}^{2}{\it a_2}\,{\it y_0}-2\,{\it y_1}\,{\it a_2}\,{
%{\it y_0}}^{2}-2\,{\it y_0}\,{\it y_1}\,{\it a_1}-{{\it y_0}}^{2}
%{\it a_2}\,{\it k_1}+{\it y_0}\,{\it a_1}\,{\it k_1}-{\it a_1}\,
%{\it k_0}\,{\it k_1}+2\,{{\it y_0}}^{2}{\it a_1}+{\it y_1}\,{\it
%a_1}\,{\it k_0}-{{\it y_1}}^{2}{\it a_2}\,{\it k_0},
%\nonumber\\
%{\it b_2}&=&2\,{\it y_0}\,{\it a_1}+{\it a_1}\,{\it k_0}-2\,{\it y_1}\,{\it
%a_1}-{\it y_1}\,{\it a_2}\,{\it k_0}-2\,{\it y_0}\,{\it y_1}\,{
%\it a_2}+{\it a_1}\,{\it k_1}+2\,{{\it y_1}}^{2}{\it a_2}-{\it
%a_2}\,{\it k_1}\,{\it k_0}-{\it y_0}\,{\it a_2}\,{\it k_1},
%\\
%{\it c_1}&=&-{\it y_0}\, ( {{\it y_1}}^{2}{\it a_2}-{\it y_1}\,{\it a_1}
%-{\it y_0}\,{\it y_1}\,{\it a_2}-{\it y_0}\,{\it a_2}\,{\it k_1}
%+{\it y_0}\,{\it a_1}+{\it a_1}\,{\it k_1} ),
%\nonumber\\
%{\it c_2}&=&-{{\it y_1}}^{2}{\it a_2}+{\it y_1}\,{\it a_1}+{\it y_0}\,{\it
%y_1}\,{\it a_2}+{\it y_0}\,{\it a_2}\,{\it k_1}-{\it y_0}\,{\it
%a_1}-{\it a_1}\,{\it k_1}.\nonumber
%\end{eqnarray*}
%}
From the representation, we need to require
\begin{equation}\label{eq-c1}-{\it y_1}+{\it k_1}+{\it y_0}\neq0,\end{equation}
and $d\,x+1$ has no roots in $[0,1]$, that is, $d>-1$, from which we can derive that
\begin{equation}\label{eq-c2}\left( -{\it y_1}+{\it k_0}+{\it y_0} \right)  \left( -{\it y_1}+{\it k_1}+{\it y_0} \right)<0.\end{equation}

From the mean value theorem, we know that $y_1-y_0=k_x$, where $k_x$ is the tangent direction of some $x\in(0,1)$. Since $C$ is monotonous, so $k_x$ is some value between $k_0$ and $k_1$. Thus conditions \bref{eq-c1} and \bref{eq-c2} are satisfied directly. We can easily transform the interval $[0,1]$ to $[x_0,x_1]$ by setting $x=\frac{X-x_0}{x_1-x_0}$, where $x\in [0,1]$ when $X\in [x_0,x_1]$. %It is clear that we can rewrite (\ref{eq-r}) as quasi rational linear form.

Furthermore, when $d\neq 0$, that is to say $-2\,{\it y_1}+{\it
k_0}+{\it k_1}+2\,{\it y_0}\neq 0$.
Then  expression \bref{eq-y1}
can be transformed into \begin{equation}\label{eq-r2}
Y_1(x)=\tilde{a}\,
x+\tilde{b}+\frac{\tilde{c}}{\tilde{d}\,x+1}.\end{equation}
%After
%simple computation, we can get
%{\tiny
%\begin{eqnarray*} \tilde{a}&=&{\frac {-{{\it y_0}}^{2}+2\,{\it y_0}\,{\it
%y_1}-{{\it y_1}}^{2}+{\it k_0} \,{\it k_1}}{{\it k_0}+2\,{\it
%y_0}-2\,{\it y_1}+{\it k_1}}}
%,\\
%\tilde{b}&=&{\frac {-2{\it k_1}{\it k_0}{\it y_1}+2{\it k_1}{\it
%y_0}{\it y_1}-{\it y_1}{{\it k_0}}^{2}+2{\it k_0}{{\it
%y_1}}^{2}+5{\it y_1} {{\it y_0}}^{2}-{\it y_0}{{\it y_1}}^{2}+{\it
%k_1}{{\it k_0}}^{2}-3{ \it k_1}{{\it y_0}}^{2}+{\it k_1}{{\it
%y_1}}^{2}-2{\it k_0}{{\it y_0 }}^{2}-{{\it y_1}}^{3}-3{{\it
%y_0}}^{3}-{\it y_0}{{\it k_1}}^{2}}{-
% \left( {\it k_0}+2{\it y_0}-2{\it y_1}+{\it k_1} \right) ^{2}}}
%,\\
%\tilde{c}&=&{\frac { \left( -{\it y_1}+{\it k_1}+{\it y_0} \right)
%\left( {{\it k_0}} ^{2}+2\,{\it k_0}\,{\it y_0}-2\,{\it k_0}\,{\it
%y_1}+{{\it y_0}}^{2}-2\,{ \it y_0}\,{\it y_1}+{{\it y_1}}^{2}
%\right) }{ \left( {\it k_0}+2\,{\it y_0 }-2\,{\it y_1}+{\it k_1}
%\right) ^{2}}}
%,\\
%\tilde{d}&=&-{ \frac {-2\,{\it y_1}+{\it k_0}+{\it k_1}+2\,{\it
%y_0}}{-{\it y_1}+{\it k_1}+{\it y_0}}}.
%\end{eqnarray*}}
Though equation \bref{eq-r2} is equivalent to equation \bref{eq-y1}
when $d\neq 0$ essentially, it has a simpler form and can reduce
computation when evaluation. When $d=0$, equation \bref{eq-y1} is a polynomial of degree two. And we
have simple expressions for parameters ${\it a,b,c}$, that is ${\it
a}=\frac{{\it k_1}-{\it k_0}}{2}$, ${\it b}=k_0$, ${\it c}=y_0$.
%This situation can easily be dealt with if we are able to early
%detection of condition $\frac{-2\,({\it y_1}-\,{\it y_0})}{{\it
%x_1}-{\it x_0}}+{\it k_0}+{\it k_1}$ equals to zero.

\noindent{\bf $C$ containing a VT point.} When a given regular curve segment contains a VT point, it means the tangent line at $P_0$ or $P_1$ is a vertical line $x-x_0=0$ or $x-x_1=0$. In this case, the method above does not work. But we can use part of an ellipse or a  hyperbola $\frac{(x-x_o)^2}{a^2}\pm\frac{(y-y_o)^2}{b^2}-1=0 (a>0, b>0)$ to derive an approximate parametrization of a real plane algebraic curve. Note that a regular curve segment containing a VT point has four cases which exactly correspond to the four parts of an ellipse or a hyperbola: the vertical line is $x-x_0=0$ or $x-x_1=0$ and $y\ge y_o$ or $y\le y_o$. We consider the case that $C$ has a vertical tangent line at $P_0(x_0,y_0)$ and $C$ monotonously increases from $P_0$ to $P_1$. And we assume that the tangent line at $P_1$ is $k_1(x-x_1)-(y-y_1)=0(k_1\ge0)$. The approximate curve is $Y_2(x)= y_o + \frac{b \sqrt{|a^2-(x-x_o)^2}|}{a}$. Note that we have $x_o=x_0\pm a, y_o=y_0$ from the property of the ellipse or the hyperbola. So we have
\begin{eqnarray}\label{eq-y2}
Y_2(x)= y_0 + \frac{b \sqrt{|a^2-(x-x_0-a)^2}|}{a}.
\end{eqnarray}
And $Y_2(x_1)=y_1,Y_2'(x_1)=k_1.$
Solving it, we have
{\small
\begin{eqnarray*}
a&=&{\frac { \left( {\it x_1}-{\it x_0} \right)  \left( {\it x_0}\,{
\it k_1}+{\it y_1}-{\it y_0}-{\it x_1}\,{\it k_1} \right) }{{\it
y_1}-{\it y_0}-2\,{\it x_1}\,{\it k_1}+2\,{\it x_0}\,{\it k_1}}},\,\,\,\\
b&=&\left( {\it x_0}\,{\it k_1}+{\it y_1}-{\it y_0}-{\it x_1}\,{\it k_1} \right) \sqrt{{\frac {{\it y_1}-{\it y_0}}{{\it y_1}-{\it
y_0}-2\,{\it x_1}\,{\it k_1}+2\,{\it x_0}\,{\it k_1}}}}\nonumber.
\end{eqnarray*}
}

From the representation, we can find that $a,b$ are well defined if $k_1<\frac{y_1-y_0}{2\,(x_1-x_0)}$ for an ellipse or $k_1>\frac{y_1-y_0}{2\,(x_1-x_0)}$ for a hyperbola. So we can choose $(x_1,y_1)$ on the regular curve segment such that $k_1\neq \frac{y_1-y_0}{2\,(x_1-x_0)}$. Then we can use part of an ellipse or a hyperbola to approximate the regular curve segments with VT points. The other three cases can be solved in a similar way.

\begin{lemma} \label{lem-in}
The two kinds of approximation curves above, say $\tilde{C}: (x,h(x)), x\in [x_0,x_1]$ for $C$ are inside the triangle formed by the endpoints and the tangent directions at the endpoints of $C$, denoted as $\Delta$.
\end{lemma}
\noindent{\bf Proof.}
For both cases, $\tilde{C}$ is part of a quadric curve. And the curve intersects all three edges of $\Delta$ at least twice (including the multiplicities). If $\tilde{C}$ goes out of $\Delta$, it will intersect the edge(s) at least three times (including multiplicities). But it is not possible since $\tilde{C}$ is part of a quadric curve. So the lemma is true.
%We prove the lemma for two cases containing VT point or not. For the fist one, if $\tilde{C}$ goes out of $\Delta$, say it goes out from the tangent line of the regular curve segment at $P_0$, then the tangent line has at least three intersection points (including the multiplicities) with $\tilde{C}$ since the approximation curve will go back to $P_1$. Note that the multiplicity of the intersection point at $P_0$ is at least two. Since $h(x)$ has form \bref{eq-r}, $\tilde{C}$, which is a quadric curve, and a line have at most two intersection points (including the multiplicities). It is a contradiction. For the tangent line of the regular curve segment at $P_1$ is samilar. For the line of the third edge of $\Delta$, the approximation curve has four intersection points with the line if $\tilde{C}$ goes out of (or tangent to) $\Delta$ from the line. It is also not possible. So $\tilde{C}$ is inside $\Delta$ for all possible $a_1,a_2$ for the first case. For the second case, it is obvious that $\tilde{C}$ is inside $\Delta$ since $\tilde{C}$ is part of a quadrics. It has two intersection points with a line at most. This ends the proof.
$\Box$

\noindent{\bf Topology preserving approximation.}
After we get the approximation regular curve segments, we need to check whether the approximation curve change the topology of the original curve. Even if we get the correct topology of the given algebraic planar curve, the approximation curve may have a different topology as the original curve, especially when two regular curve segments are very close, for example, see Figure 2. So we need to ensure that our numeric approximation curve has the same topology as the original one. We need only ensure that any two approximation curves, say $C_1,C_2: (x,p(x)), (x,q(x)),x\in[a,b]$, are disjoint.
If $p(x)-q(x)=0$ has no real roots in $(a,b)$, then the two approximation regular curve segments are disjoint. There are two kinds of approximation curves, say $Y_1(x), Y_2(x)$. So we need to consider:\\
\noindent Case one: two approximation curves are both rational ones as $Y_1(x)$. Then $T(x)=p(x)-q(x)$ can be simplified into a cubic univariate polynomial. It is easy to check whether it contains a real roots in $(a,b)$ by its coefficients.\\
Case two: one is as $Y_1(x)$ and the other is as $Y_2(x)$. Then $T(x)=p(x)-q(x)$ can be simplified into a quartic univariate polynomial. It is also easy to check whether it contains a real roots in $(a,b)$ by its coefficients.\\
Case three: both approximation curves are as $Y_2(x)$. They both are parts of quadric algebraic curves. Considering the intersection of two quadric algebraic curves, we can judge whether $C_1,C_2$ are disjoint or not.

Doing so as above, our approximation is exactly topology preserving.

\subsection{Error control of the approximation}
We will show the error control of the plane approximation curve in this subsection.
In geometry, the approximation error should be defined as the following Hausdorff distance between
the segment $S$ and its approximation $S_a$,
\begin{equation}\label{eq-herr}
e(S, S_a) = dis(S, S_a) = \max_{P\in S}\min_{P'\in S_a} d(P,P').
\end{equation}
However such a distance is difficult to compute. As an
implement, the distance from an approximation parametric curve $P(t) = (x(t), y(t)), 0 \le t \le 1$ to the implicit
defined curve $\Cc : f (x, y) = 0$ is taken in the following form, which is called the error function \cite{chuang},
\begin{equation}\label{eq-err}
e(t) = \frac{f (x(t), y(t))}{\sqrt{(f_x(x(t), y(t))^2 +f_y(x(t), y(t))^2}}.
\end{equation}
The approximation error between $P(t)$ and $\Cc$ is set as an optimization problem $$e(P(t), \Cc) = \max_{0\le t\le 1}(e(t)).$$

%For our approximation, we use another way.
 Let $C: (x, \tilde{y}(x)), x\in[x_0,x_1]$ be the regular curve segment and $\overline{C}:=(x, Y(x)), x\in[x_0,x_1]$ its approximation curve. It is not difficult to find that the following bound is an upper bound of the Hausdorff distance between
the segment $C$ and its approximation curve $\overline{C}$ from (\ref{eq-herr}):
\begin{equation}\label{eq-myerr}
\max_{x\in[x_0,x_1]}|Y(x)-\tilde{y}(x)|.
\end{equation}

We use Newton-Ralphson method to obtain $\tilde{y}(x_i^0)$ at some point $x_i^0\in [x_0,x_1]$ in practice and $Y(x_i^0)$ is the start point. If we fail to get a point with  Newton-Ralphson method or the point satisfying $|Y(x_i^0)-\tilde{y}(x_i^0)|\ge \delta$, we can divide the regular curve segment into two ones. The approximation error is bounded by $\max_i\{|Y(x_i^0)-\tilde{y}(x_i^0)|\}$. In practice, we sample $x_i^0$ as $x_i^0 = x_0+i/n\,(x_1-x_0), 0 \le i \le n$, for a proper value of $n$.

%$$e(x) = \frac{f (x, Y(x))}{(f_x(x, Y(x))^2 +f_y(x, Y(x))^2)^{\frac{1}{2}}}. $$

In order to control the error under a given precision, we need to divide the regular curve segment into two or more regular curve segments recursively until the error requirement satisfied. We subdivide the regular curve segments into two or more regular curve segments uniformly in the $x$ coordinate. For any regular curve segment $C: (x, \tilde{y}(x)),x\in[a,b]$, we denote the endpoints as $P_0(x_0,y_0),P_1(x_1,y_1)$ and the tangent directions as $(1,k_i), i=0,1$. We can find that $P_0,P_1$ and two tangent directions form a triangle.
One can subdivide the regular curve segment into two (or more) ones, for example, $C_1: x\in[x_0,(x_0+x_1)/2],C_2: x\in[(x_0+x_1)/2,x_1]$, if the precision is not satisfied. For the approximation curve of our method, we will prove that it can achieve any given precision.

\begin{theorem}\label{thm-err}
Let $P_0,P_1$ be the endpoints of a regular curve segment $C$ and $\Delta$ the triangle related to the regular curve segment as defined before. The Hausdorff distance between $C$ and its approximation curve(s) tends to zero if we subdivide $C$ into two or more regular curve segments recursively.
\end{theorem}
\noindent{\bf Proof.}
We will consider divide $C$ into two regular curve segments for the proof since dividing them into more regular curve segments are similar. Let $P((x_0+x_1)/2,\bar{y})$ be a point on $C$. Denote the triangles formed by $P_1,P(P,P_2)$ and the tangent directions of $C$ at these points as $\Delta_1(\Delta_2)$. Let the lengths of the line segments $\overline{P_1P},\overline{PP_2}$ be $L_1^1, L_2^1$ and the heights of the triangles $\Delta_1,\Delta_2$ corresponding to the edges $\overline{P_1P},\overline{PP_2}$ are $H_1^1,H_2^1$, as shown in Figure 3. Subdividing the regular curve segments recursively in a similar way, denoting the length of the edges and heights as $L_j^i, H_j^i$ of the triangles, we have the sum of the areas of these triangles are
$$A=\sum_j (L_j^i\,H_j^i/2)<\frac{1}{2}\sum_j L_j^i\max_j H_j^i.$$
Assume that one edge $L_j^i=\overline{P_1'P_2'}, P_1'(x_1',y_1'),P_2'(x_2',y_2')$. Since the given regular curve segment is bounded, $|y_2'-y_1'|$ tends to zero when $|x_2'-x_1'|$ tends to zero. And $\sum_j L_j^i$ tends to the length of arc, say $\mathcal{L}$, of the given regular curve segment and $H_j^i$ tends to zero when all corresponding $|x_2'-x_1'|$ of $L_j^i$ tends to zero. Thus $A$ tends to zero. From the result of Lemma \ref{lem-in}, we can have the opinion that we have a proof of the theorem.
$\Box$

There are two ways to find the subdivision points on a given regular curve segment $C: (x, \tilde{y}(x)),x\in [a,b]$. Since we get the topology of the plane projection curve, we know the order of the given regular curve segment among all the regular curve segments of the projection curve $h(x,y)=0$ when $x$ changes from $a$ to $b$. That is, we can find the point on $C$ for a fixed $x$ coordinate, say $x_0\in(a,b)$. It is the real root with the same order of $h(x_0,y)=0$ in a fixed interval (or $(-\infty,+\infty)$).

Another way is a local method. We can trace the regular curve segment to find the point on $C$ with given $x$ coordinate since the regular curve segments are monotonous and convex. From the endpoint of the regular curve segment, compute the tangent line of the regular curve segment at some point $P$, find a point $Q$ on the tangent line by increasing the $x$ coordinate such that the line segment $\overline{PQ}$ has no intersection with the projection curve. Then fix the $x$ coordinate of $Q$, to find a point on the regular curve segment. Note that we know the direction to find the point from the positiveness or negativeness of the tangent direction. Doing so recursively, we can find the point that we want, as shown in Figure 4.

    \begin{figure}[hb]
\begin{minipage}{0.234\textwidth}\label{fig-err-app}
\centering
\includegraphics[height=25mm]{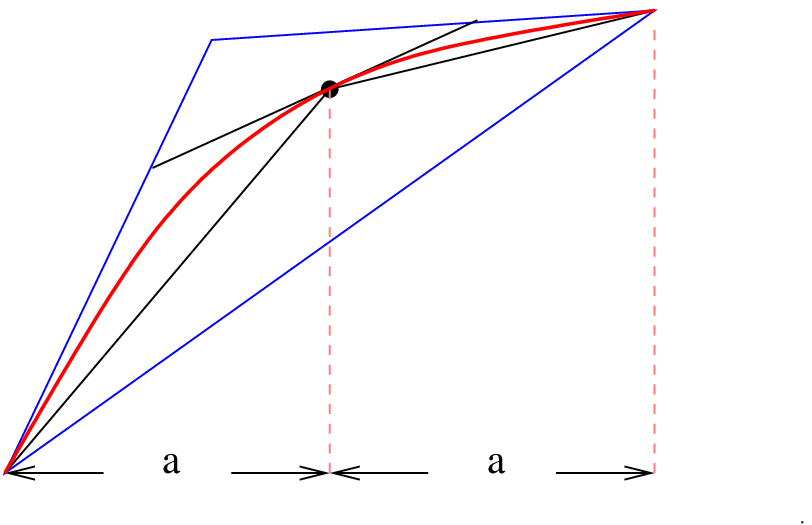}
\caption{Splitting a regular curve segment}
\end{minipage}
\begin{minipage}{0.234\textwidth}\label{fig-err-app}
\centering
\includegraphics[height=25mm]{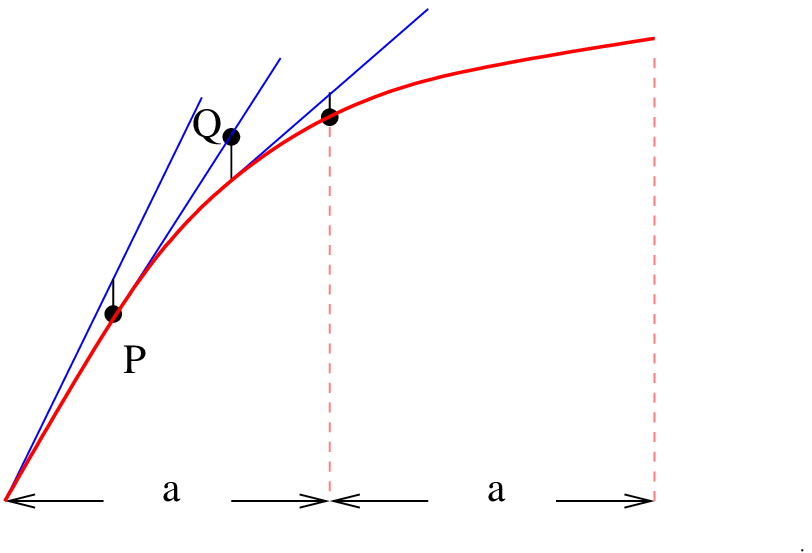}
\caption{Finding subdivision point by tracing}
\end{minipage}
    \end{figure}

%\subsection{Main algorithm for approximate parametrization of real algebraic curves}
With the preparation above, we have the following algorithm to approximate a plane algebraic curve.
\begin{algo} The inputs are $\Cc : f (x, y) = 0$, a bounding box
$B$ and an error bound $\delta >0$. The outputs are parametric curves $\Cc_1:=\{B_i(x)=(x,y_i(x)),a_i\le x\le b_i, (i=1,...,N)\}$, such that they give a $C^1$-continuous and topology preserving approximation to $\Cc_B$ with $e(\Cc_1, \Cc) < \delta$.
\end{algo}
\begin{enumerate}
\item Regular curve segmentation of $\Cc_B$ as in Section 2.2.
\item Regular curve segment approximation as in Section 2.3 with error control of the approximation as in Section 2.4.
%\item Topology determination of $\Cc_B$.
%%Determine the bounding box $\B$ and the topology of $\CB$.
%%Let the resulted segments be $ST$ and let $GT = U(ST )$.
%\item Flexes computation and $x$-critical points computation. Divide $\Cc_B$ into regular curve segments such that $x$($y$)-critical points and flexes appear only at the endpoints of these regular curve segments.
%\item Tangent computation. Compute the (approximation) tangent directions of at each endpoints of the regular curve segment.
%\item Segment approximation. Approximate each regular curve segment with explicit parametric curves under the given precision $\delta$ by dividing the regular curve segments into two or more regular curve segments recursively. If two regular curve segments are not disjoint, subdivide the original regular curve segments and approximate again until the approximation regular curve segments are disjoint and satisfies the precision.
\end{enumerate}
 The correctness of the algorithm is clear from the analysis above. The termination of the algorithm is guaranteed by Lemma \ref{lem-in} and Theorem \ref{thm-err}.

\section{Certified approximate parametrization of algebraic space curves}
In this section, we will consider approximate parametrization of algebraic space curves defined by $f,g\in \Q[x,y,z]$ such that
two assumptions hold:
\begin{itemize}
\item For any $x_0\in\R$, $f(x_0,y,z)=g(x_0,y,z)=0$ has a finite number of solutions; and
\item the leading coefficients of $f,g$ w.r.t. $z$ have no common factors only in $x$.
\end{itemize}
The assumptions are to ensure that we can use local generic position
method to recover the points on the space curve from the points on
two plane projection curves. The first assumption ensures that the
algebraic space curve defined by $f=g=0$ does not have a plane curve
on the plane $x=x_0$.
%Thus on the fiber $x=x_0$, the two projection curves have only finite points.
%Then each point on $h=0$ corresponds to finite point of the algebraic space curve.
The second assumption ensures that we can find a generic position
only by taking a shear map $(x,y,z)\rightarrow
(x,y+s\,z,z)$. If the two projection curves have no factors only
involving $x$, the two assumptions hold.

%The method also has limitations: when the projection planar curves
%$h(x,y)=0, \bar{h}(x,y)=0$ have vertical lines, that is, the
%defining polynomial for the projection curve has factor(s) involving
%only $x$, we can not recover the $z$-coordinate of the space curve.
%Thus we have two {\bf assumptions} for the input polynomials. First,
%for any $x_0\in \R$, $f(x_0,y,z),g(x_0,y,z)$ have finitely many
%solutions. Second, the gcd of the leading coefficients of $f, g$
%w.r.t. $z$ has no factors involving only $x$.
In fact, most of the
problems we considered satisfy the condition. Note that we can
exchange $x,y,z$ freely. And another coordinate system transformation $(x,y,z)\rightarrow (x+s\,z,y,z)$ can help us to find out the missing regular curve segments in the first transformation, even when the algebraic space curve containing vertical lines. Thus we can remove the assumptions with the method mentioned here. But we still assume the two assumptions holds in this section.

\subsection{Definition of local generic position}
In order to reduce the 3D approximation of space curves into 2D
approximation of plane curves, we need the concept of local generic
position. We recall the related definitions for zero-dimensional
bivariate polynomial system \cite{lgp}. Let $ \C$ be the field
of complex numbers. Let $f,g\in \Q[x,y]$.
    \begin{figure}[hb]
    \centering
\begin{minipage}{0.480\textwidth}\label{fig-lgp}
\centering
\includegraphics[height=48mm]{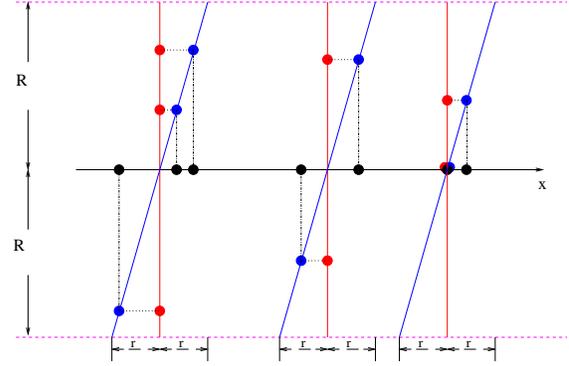}\caption{LGP method}
\end{minipage}
\end{figure}
We say two plane curves defined by two polynomials $f,g$ such that $\gcd(f,g)=1$ are in a {\bf generic position} w.r.t. $y$ if\\
%\indent 1) $\gcd(f,g)=1$.\\
\indent 1) The leading coefficients of $f$ and $g$ w.r.t. $y$ have no common factors.\\
\indent 2) Let $h$ be the resultant of $f$ and $g$ w.r.t. $y$. For any $\alpha\in \C$ such that $h(\alpha)=0$, $f(\alpha,y),g(\alpha,y)$
 have only one common zero in $\C$.

%Since we consider only the real part of the algebraic space curves, we give the following definition.
%
%Two polynomial $f,g\in K[x,y]$ are called in a {\bf real generic position} w.r.t. $y$ if\\
%\indent 1) $f,g$ are in a generic position w.r.t. $y$.\\
%\indent 2) For any $\alpha\in \R$ such that $h(\alpha)=0$, $f(\alpha,y),g(\alpha,y)$ have and only have one distinct common real zero
% in $\R$.

Then we will introduce the technique of local generic position (LGP for short) method.

Given $f,g\in \Q[x,y]$, not necessarily to be in generic position, we can take a coordinate system transformation $\phi: (x,y)\rightarrow(x+s\,y,y), s\in\Q$ such that
\begin{itemize}
\item $\phi(f),\phi(g)$ are in a generic position w.r.t. $x$.
%\item Let $\bar{h}$ (or $h$) be the resultant of $\phi(f),\phi(g)$ (or $f,g$) w.r.t. $y$. For any $\alpha\in \R$ such that $h(\alpha)=0$, it has a length-fixed neighborhood $H_\alpha$ such that $H_\alpha\cap H_\beta=\emptyset$ for any distinct real roots $\alpha,\beta$ of $h=0$. Any root $(\gamma,\eta)$ of $f=g=0$ which has a same $x$-coordinate $\gamma$, is mapped to $\gamma'=\gamma+s\,\eta\in H_\gamma$, where $h(\gamma)=0, \bar{h}(\gamma')=0$, as shown in Figure 3. Thus we can recover $\eta=\frac{\gamma'-\gamma}{s}$. And the error for the coordinate $\eta$ is $\frac{\epsilon_1+\epsilon_2}{s}$ if the errors for $\gamma, \gamma'$ are $\epsilon_1,\epsilon_2$, respectively.
\item Let $\bar{h}$, $h$ be the resultants of $\phi(f),\phi(g)$ and $f,g$ w.r.t. $y$, respectively. Each root $\alpha$ of $h(x)=0$ has a neighbor interval $H_\alpha$ such that $H_\alpha\cap H_\beta=\emptyset$ for roots $\beta\neq\alpha$ of $h=0$.
And any root $(\gamma,\eta)$ of $f=g=0$ which has a same $x$-coordinate $\gamma$, is mapped to $\gamma'=\gamma+s\,\eta\in H_\gamma$, where $h(\gamma)=0, \bar{h}(\gamma')=0$, as shown in Figure 5.
%\ref{fig-lgp}.
Thus we can recover $\eta=\frac{\gamma'-\gamma}{s}$.
\end{itemize}
We can find the method has two nice properties: 1) The 2D solving problem is transformed into a 1D solving problem. 2) The error control of the solutions is easier.

\subsection{Basic idea}
Now we want to extend this technique to 3-D case.
Let $f \wedge g $ denote the algebraic space curve defined by $f=g=0$. Denote $\pi_z: (x,y,z)\rightarrow (x,y)$ and $h=\pi_z(f\wedge g)$. Let $\varphi: (x,y,z)\rightarrow (x,y+s\,z,z)$,
$\bar{h}=\pi_z(\varphi(f)\wedge\varphi(g))$.  For a proper regular curve segment $C: (x,p(x)),x\in[x_0,x_1]$ of the plane curve defined by $h=0$, it corresponds to one (the corresponding space regular curve segment may be at infinity) or more space regular curve segment(s), denoted as $S_1,\ldots,S_t$. If we can choose a proper $s$ such that $\varphi(f), \varphi(g)$ is in ``a good'' position, and it has some local property, that is, the corresponding projection regular curve segments of $S_1,\ldots,S_t$, say $C_i: (x,q_i(x)),x\in[x_0,x_1](i=1,\ldots,t)$, are in a fixed neighborhood of $C$, then we can recover the $z$-coordinate of $S_i: z=\frac{q_i(x)-p(x)}{s}(i=1,\ldots,t)$. But to make $C_i$ in a fixed neighborhood of $C$ is an unreachable task sometimes. Fortunately, what we need is to find out the correspondence between the regular curve segments of the plane curves defined by $h=0$ and $\bar{h}=0$. We can choose some sample points on $x$-axis, say $x_i^{(0)}, i=1,\ldots,n$, a proper $s$ such that $\varphi(f(x_i^{(0)},y,z)),\varphi(g(x_i^{(0)},y,z))$ are in a generic position. Then we can figure out the correspondence between the regular curve segments of $h=0$ and $\bar{h}=0$, as shown in Figure 6.
%The red ones are $S_i(i=1,2,3)$ and the black one is their projection in $h=0$. The blue ones are the corresponding regular curve segments of $S_i(i=1,2,3)$ in $\varphi(f)\wedge\varphi(g)$ and the gray ones are their projections $C_i(i=1,2,3)$ in $\bar{h}=0$.

%Given two algebraic surfaces defined by $f,g\in \Q[x,y,z]$ such that two assumptions hold:
%\begin{itemize}
%\item For any $x_0\in\R$, $f(x_0,y,z)=g(x_0,y,z)=0$ has a finite number of solutions; and
%\item the leading coefficients of $f,g$ w.r.t. $z$ have no common factors only in $x$.
%\end{itemize}
%The assumptions are to ensure that we can use local generic position method to recover the points on the space curve from the points on two plane projection curves.
%The first assumption ensures that the algebraic space curve defined by $f=g=0$ does not have a plane curve on the plane $x=x_0$. Thus on the fiber $x=x_0$, the two projection curves have only finite points.
%%Then each point on $h=0$ corresponds to finite point of the algebraic space curve.
%The second assumption ensures that we can find a generic position only by taking coordinate system transformation $(x,y,z)\rightarrow (x,y+s\,z,z)$. If the two projection curves have no factors only involving $x$, the two assumptions hold.

To realize the aim above, there are two key steps. One is how to find an approximation $s$, the other is how to find the correspondence between the regular curve segments of $h=0$ and $\bar{h}=0$. We need some preparations at first.

We say that two algebraic surfaces defined by $f,g\in \Q[x,y,z]$ such that $\gcd(f,g)=1$ are in a {\bf $z$-generic position} if\\
%\indent 1) $\gcd(f,g)=1$.\\
\indent 1) The leading coefficients of $f$ and $g$ w.r.t. $z$ have no common factors.\\
\indent 2)  Let $h$ be the resultant of $f$ and $g$ w.r.t. $z$. There are only a finite number of zeros $(\alpha,\beta)\in \C^2$ such that $(\alpha,\beta)$ is not a $y$-critical point, $h(\alpha,\beta)=0$ and $f(\alpha,\beta,z),g(\alpha,\beta,z)$ have more than one distinct common zeros in $\C$.

The definition is similar to the definition of pseudo-generic position in \cite{dao}.
%Our definition is a special case of pseudo-generic position.
In Theorem 4 of \cite{dao}, the authors also provide a method to check whether two given surfaces are in a pseudo-generic position or not.

\noindent{\bf Computing $s$.} We will show how to find an $s$ mentioned before. Let $\pi_y: (x,y)\rightarrow (x)$. Denote the real roots of $\pi_y(h)=0$ and the $x$-coordinates of the flexes and $x$-critical points of $h=0$ as $\alpha_1,\alpha_3,\ldots,\alpha_{2\,t-1}$. Find two rational numbers less than $\alpha_1$ and larger than $\alpha_{2\,t-1}$, denoted as $\alpha_0, \alpha_{2\,t}$ respectively. For any two adjacent real roots $\alpha_{2\,i-1}, \alpha_{2\,i+1}$ of $\pi_y(h)=0$, we can find a rational number, say $\alpha_{2\,i}$. Then we obtain a sequence $\alpha_i (i=0,\ldots,2\,t)$. Assume the real roots of $h(\alpha_i,y)=0$ are $\beta_{i,j} (j=0, \ldots,t_i)$ which are listed in increasing order. We can find out that $(\alpha_i,\beta_{i,j})$ divide the plane curve $h=0$ in the region $[\alpha_0,\alpha_{2t}]\times \R$ into regular curve segments. Let
{\tiny \begin{eqnarray}
R&=&\max_{0\le i\le 2\,t,0\le j\le t_i}{\rm RB_z}(f(\alpha_i,\beta_{i,j},z)),\nonumber\\
r&=&\min_{0\le i\le 2\,t}\{R,\min_{0\le j\le t_i-1}(\beta_{i,j+1}-\beta_{i,j})\},\nonumber\\
0&<&s<\frac{r}{2\,R}, s\in\Q,   \label{eq-s}
\end{eqnarray}
}
 where $\beta_{i,-1}=-\infty$, ${\rm RB_z}(f(\alpha_i,\beta_{i,j},z))$ is the root bound of $f(\alpha_i,\beta_{i,j},z)$ in $z$, $f$ can be replaced by $g$.  Since it is probability 1 under condition (\ref{eq-s}) to obtain such an $s$ that $\varphi(f(\alpha_i,y,z)), \varphi(g(\alpha_i,y,z))$ are in a generic position with the assumptions, it is probability 1 that $\varphi(f(\alpha_i,y,z)), \varphi(g(\alpha_i,y,z))$ are in a local generic position for all $\alpha_i (i=0,\ldots,2\,t)$.
 And we can ensure this by checking whether $\varphi(f)\wedge \varphi(g)$ is in a $z$-generic position.

 \noindent{\bf Finding the correspondence.}
 With local generic position method and the assumptions, we can recover the points of $f\wedge g$ corresponding to $(\alpha_i,\beta_{i,j})$, say $(\alpha_i,\beta_{i,j},\gamma_{i,j,k}) (1\le k\le t_{i,j})$. As shown in Figure 6, from $B(\alpha,\beta), B_1(\alpha,\beta_1)$, we can find out the point corresponding to $B,B_1$ in 3D space: $(\alpha,\beta,\frac{\beta_1-\beta}{s})$. Note that $B_1$ is in a neighborhood $\alpha\times(\beta-r/2,\beta-r/2)$ of $B$.

Let $\alpha,\alpha'$ be any $\alpha_i,\alpha_{i+1}$.  %for $i=0,\ldots,2t-1$.
We will classify the piece of curves inside $(\alpha,\alpha')\times\R$ into two cases by considering whether they contain singularities.

If the two endpoints of a regular curve segment $C$ of $\bar{h}=0$ are in the fixed neighborhood of the endpoints of a regular curve segment $\bar{C}$ of $h=0$ respectively, we know that $\bar{C}$ corresponds to $C$, see $\widetilde{R_iU_i} (i=1,2)$ and $\widetilde{RU}$ in Figure 6 for example.

 There are two cases for the singular points of $\bar{h}=\pi_z(\varphi(f)\wedge\varphi(g))=0$ in $(\alpha,\alpha')\times\R$: One case is that some correspond to singularities of $f\wedge g$. Two or more space regular curve segments of $f\wedge g$ intersect on a cylinder surface defined by some factor(s) of $h=0$. So this kind of singularities of $f\wedge g$  may not correspond to singularities of $h=0$. If two or more left (right) branches of a singularity of $\bar{h}=0$ correspond to a same regular curve segment of $h=0$, we can judge that it is a true singularity of $f\wedge g$, see the point $D$ in Figure 6 for example. The regular curve segments $\widetilde{A_1DL_2},\widetilde{A_2DL_1},\widetilde{A_3EFL_3}$ belonging to $\bar{h}=0$ all correspond to $\widetilde{AR}$ since $A_1,A_2,A_3(R_1,R_2,R_3)$ are in a neighborhood of $A(R)$.
 %Note that we can get the information when we compute the topology of $\bar{h}=0$ with some property of local generic position.

The other case is that they are not true singularities of $f\wedge g$.
Thus the curve branches of $\bar{h}=0$ which pass through these singular points of $\bar{h}=0$ with different tangent lines correspond to disjoint space curves of $\varphi(f)\wedge \varphi(g)$, then we can find out the correspondence of $C$ and $C_i$, see points $E,F$ in Figure 6 for example.  Note that the continuous space curve maps to a continuous plane curve by $\pi_z$. If there exist two or more curve branches having same tangent lines at a singular point of $\bar{h}=0$ in $(\alpha,\alpha')\times\R$, see point $G$ in Figure 6 for example, we call it {\bf tangent false singularity}. If $C_i$ contains only one tangent false singularity in $(\alpha,\alpha')$, we can still find out the correspondence of $C_i$ and its corresponding regular curve segment $C$ following the correspondence of the endpoints of $C$ and $C_i$. Note that if the endpoints of two regular curve segments of $f\wedge g$ have same $x,y$ coordinates, their corresponding projection curves overlap in $h=0$, and the projection curve of their corresponding regular curve segment in $\varphi(f)\wedge\varphi(g)$ are disjoint in $\bar{h}=0$ (except the endpoints). For $G$ in Figure 6, we know the endpoints of $\widetilde{R_3GU_3}$ are in the fixed neighborhood of $R$ and $U$, so $\widetilde{R_3G},\widetilde{GU_3}$ correspond to $\widetilde{RU}$. But if $C_i$ contains two or more tangent false singularities in $(\alpha,\alpha')$, we can not determine the correspondence of the part(s) of $C_i$ between these tangent false singularities and $C$ (or other regular curve segment of $h=0$). As shown in Figure 6, $H,K$ are two tangent false singularities and we do not know the correspondence of the two regular curve segments between them. We can find back the correspondence in the following way. Let $(p_0,p_1)$ and $(q_0,q_1)$ be two adjacent tangent false singularities on $C_i$ in $(\alpha,\alpha')$. Choose a rational number $\gamma$ such that $p_0<\gamma<q_0$. Solving $f(\gamma,y,z)=g(\gamma,y,z)=0$, we can get some real points on $f\wedge g$.  Solving $\bar{h}(\gamma,y)=0$, we can get some real points on $\bar{h}=0$. Since $\varphi(f)\wedge \varphi(g)$ is in a $z$-generic position, $\varphi(f(\gamma,y,z))\wedge\varphi(g(\gamma,y,z))$ is in a generic position. So two group of points have a one-to-one map. Thus we can find out the correspondence between them. So we can decide the correspondence between $C$ and (parts of) $C_i$.

    \begin{figure}[hb]
%\begin{minipage}{0.30\textwidth}\label{fig-lgp}
%\centering
%\includegraphics[height=28mm]{fig/slope.eps}\caption{LGP method}
%\end{minipage}

%\begin{minipage}{0.24\textwidth}\label{fig-trans}
%\centering
%\includegraphics[height=28mm]{fig/trans3.eps}
%\caption{LGP in 3D space}
%\end{minipage}
\begin{minipage}{0.48\textwidth}\label{fig-tangent}
\centering
\includegraphics[height=48mm]{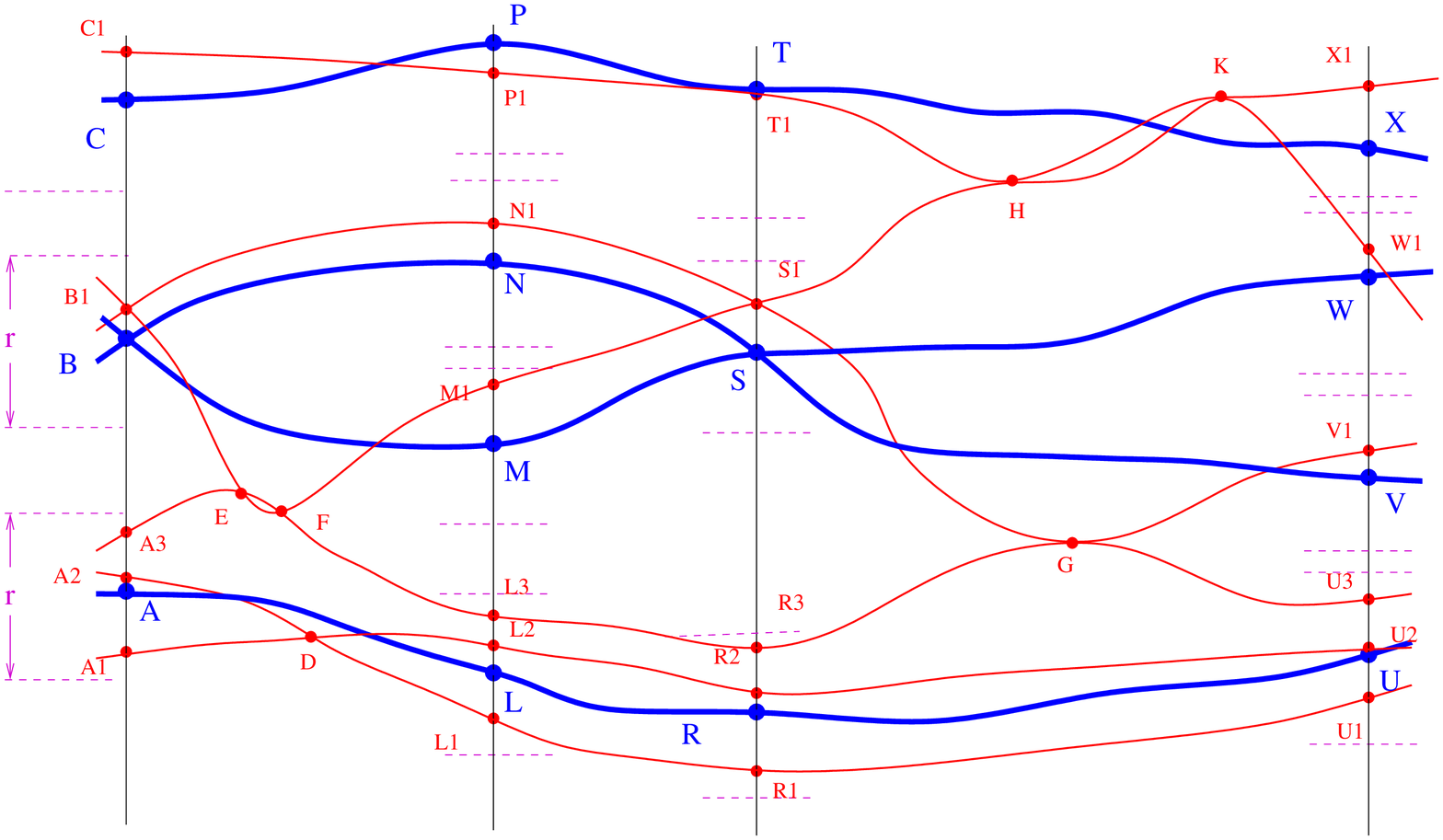}\caption{Two projection curves and their correspondences}
\end{minipage}
%\begin{minipage}{0.32\textwidth}\label{fig-example1}
%\centering
%\includegraphics[height=50mm]{fig/projection.eps}\caption{$\Cc_1, \Cc_2$ in Example 1}
%\end{minipage}
    \end{figure}

Based on the analysis above, we can write a theorem below which tell us an algorithm to obtain the correspondence between the two projection curves $h=0$ and $\bar{h}=0$. %Furthermore, the algorithm computes the topology of the algebraic space curve.

\begin{theorem}\label{thm-correspondence}
Use the same notations as above. There exists an algorithm to find a proper $s$ to obtain $\bar{h}=0$, and to find out the correspondence between the regular curve segments of $h=0$ and $\bar{h}=0$.
\end{theorem}
\noindent{\bf Some remarks for the theorem: }
\begin{enumerate}
\item  If we consider the algebraic space curve inside a box $B=[a_1,b_1]\times[a_2,b_2]\times[a_3,b_3]$,  we can consider only the points, including the $x$($y$)-critical points and flexes inside $B$. We also need to decide the intersection between the space curve and the boundaries of the box. We can use two planes $z-a_3=0, z-b_3=0$ to intersect $f, g$ and project the intersection into $xy$-plane, respectively. Replace $h$ by the product of these projections and $h$. Use two lines $y-a_2=0, y-b_2=0$ to intersect $h=0$, we get some of the projections of the boundaries of the algebraic space curve inside $B$.
 \item In fact, this theorem gives an algorithm to compute the topology of an algebraic space curve.
 %If we use line segments to replace the regular curve segments in plane, and represent the 3D line segments formed by two plane line segments, we can obtain a 3D graph which presents the topology of the algebraic space curve.
 \item In the computation in practice, $(\alpha_i,\beta_{i,j})$ are represented by isolating intervals. The corresponding method to get $r,R$ can be found in \cite{lgp}.
\end{enumerate}

If we approximate the two plane projection curves with the forms \bref{eq-y1}, \bref{eq-y2}, we can get the piecewise approximation parametric space curves of $f\wedge g$ from the correspondence of the regular curve segments in the two plane curves in Theorem \ref{thm-correspondence}. They have the form:
\begin{equation}\label{eq-app}(x,y_1(x),\frac{y_2(x)-y_1(x)}{s}), x\in [a,b]\end{equation}
where $y_1(x),y_2(x)$ are as Forms \bref{eq-y1}, \bref{eq-y2}.

We need also check whether our approximation space curve changes the topology of original space curve or not. Since the plane approximation curve does not change the topology of the plane projection curve, we need only to check whether two approximation space regular curve segments having the same $y$ coordinate are disjoint or not. We assume that the two approximation space regular curve segments are
{\small $$C_1: (x,y(x),\frac{y_2(x)-y(x)}{s}), C_2: (x,y(x),\frac{y_1(x)-y(x)}{s}), x\in[a,b].$$}
They are disjoint if $\frac{y_2(x)-y(x)}{s}-\frac{y_1(x)-y(x)}{s}=0$ has no real roots in $(a,b)$. We use the similar method as we check two plane approximation regular curve segments in Section 2.3.

\subsection{Error control of the approximation space curves}
In this subsection, we will consider how to control the error of the approximation space curve.
\begin{theorem} \label{lem-err}
Use the notations as before. If we approximate the plane curves $h=0$ and $\bar{h}=0$ with errors $\epsilon_1,\epsilon_2$, respectively, the error of each coordinate of the approximating curve of the algebraic space curve $f\wedge g$ is bounded by $\max(\epsilon_1,\frac{\epsilon_1+\epsilon_2}{s})$, and the Housdorff distance error of the approximating curve is bounded by $\frac{\sqrt{s^2\epsilon_1^2+(\epsilon_1+\epsilon_2)^2}}{s}$.
\end{theorem}
\noindent{\bf Proof.} Let $C: (x,\tilde{y}(x))$ ($C_i: (x,\tilde{y_i}(x))$) be the regular curve segment of $h=0$ ($\bar{h}=0$) and $\bar{C}: (x,p(x))$ ($\bar{C_i}: (x,q(x))$) its approximation curve, $x\in [x_0,x_1]$. $C_i$ corresponds to $C$. Let $S_i: (x, \tilde{y}(x),\tilde{z}(x))$ (exact representation) and $\bar{S_i}: (x,p(x),\frac{q(x)-p(x)}{s}),x\in[x_0,x_1]$ be a space regular curve segment and its approximation. From the condition, we have
%$$e(C,\bar{C})=\max_{P\in C}\min_{P'\in \bar{C}} d(P,P')<\epsilon_1,$$
%$$e(C_i,\bar{C_i})=\max_{P\in C_i}\min_{P'\in \bar{C_i}} d(P,P')<\epsilon_2.$$
$e(C,\bar{C})<\epsilon_1, e(C_i,\bar{C_i})<\epsilon_2.$
The error here is defined by (\ref{eq-myerr}).
Let us consider the three coordinates of one part of the approximation curve $S: (x,p(x),\frac{q(x)-p(x)}{s}),x\in[x_0,x_1]$.

The errors of the first and second coordinates are $0$ and $\epsilon_1$, respectively. For the third coordinate, we have $\tilde{z}(x)=\frac{\tilde{y_i}(x)-\tilde{y}(x)}{s}$. Thus
{\tiny \begin{eqnarray*}
&&|\tilde{z}(x)-\frac{q(x)-p(x)}{s}|=|\frac{\tilde{y_i}(x)-\tilde{y}(x)}{s}-\frac{q(x)-p(x)}{s}|\\
&\le&\frac{|\tilde{y_i}(x)-q(x)|+|\tilde{y}(x)-p(x)|}{s}\le\frac{\epsilon_1+\epsilon_2}{s}.
\end{eqnarray*}
}
So the third coordinate is bounded by $\frac{\epsilon_1+\epsilon_2}{s}$ from (\ref{eq-myerr}). %So we prove the lemma.
From the definition of Hausdorff distance (\ref{eq-herr}), we have the Hausdorff distance of $S_i$ and $\tilde{S_i}$:
{\tiny
\begin{eqnarray*}&&e(S_i,\tilde{S_i})=\max_{P\in S_i}\min_{P'\in \tilde{S_i}} d(P,P')\le \max_{P\in S_i,P'\in \tilde{S_i},P'_x=P_x} d(P,P')\\
&<&\sqrt{\epsilon_1^2+((\epsilon_1+\epsilon_2)/s)^2}=\frac{\sqrt{s^2\epsilon_1^2+(\epsilon_1+\epsilon_2)^2}}{s}.
\end{eqnarray*}
}
This ends the proof.
$\Box$

%The following corollary is clear.
%\begin{corollary}\label{cor-err}
%With the notations of Lemma \ref{lem-err}, the Housdorff distance error of the approximating curve is bounded by $\frac{\sqrt{s^2\epsilon_1^2+(\epsilon_1+\epsilon_2)^2}}{s}$.
%\end{corollary}
%\noindent{\bf Proof.}
%From the definition of Hausdorff distance (\ref{eq-herr}), we have the Hausdorff distance of $S_i$ and $\tilde{S_i}$:
%{\small
%\begin{eqnarray*}e(S_i,\tilde{S_i})=\max_{P\in S_i}\min_{P'\in \tilde{S_i}} d(P,P')\le \max_{P\in S_i,P'\in \tilde{S_i},P'_x=P_x} d(P,P')
%<\sqrt{\epsilon_1^2+((\epsilon_1+\epsilon_2)/s)^2}=\frac{\sqrt{s^2\epsilon_1^2+(\epsilon_1+\epsilon_2)^2}}{s}.
%\end{eqnarray*}
%}
%This ends the proof.
%$\Box$

If the required precision for the approximation curve is $\epsilon$, we can approximate the plane algebraic curves $h=0$ and $\bar{h}=0$ with precision $\frac{s}{\sqrt{s^2+4}}\epsilon$ from the theorem.

\subsection{$G^1$-continuous rational approximation space curve}
We will derive approximation space curve from plane approximation curve.
%We can approximate plane algebraic curves with $C^1$-continuous piecewise parametric curves.
And we will re-parameterize the non-rational parametric curve into rational ones. Thus the obtained approximation space parametric curves are $G^1$-continuous and rational.
\begin{lemma}\label{lem-continuous}
Use the notations as before. If we approximate the plane curves $h=0$ and $\bar{h}=0$ with $C^1$-continuous parametric curve, the approximation curve of the algebraic space curve $f\wedge g$ is $C^1$-continuous.
\end{lemma}
\noindent{\bf Proof.} Let $(x,p(x)),(x,q(x)),x\in [x_0,x_1]$ be two corresponding approximation curves of the regular curve segments of $h=0$ and $\bar{h}=0$ and $p(x),q(x)$ are $C^1$-continuous in $[x_0,x_1]$. We can obtain the approximating curve of the space regular curve segment: $S: (x,p(x),\frac{q(x)-p(x)}{s}),x\in[x_0,x_1]$. The tangent direction of $S$ at any $x$ is $(1,\frac{\partial p}{\partial x},(\frac{\partial q}{\partial x}-\frac{\partial p}{\partial x})/s)$. From the definition of $C^1$-continuous, we can find that $S$ is $C^1$-continuous since $(x,p(x)),(x,q(x)),x\in [x_0,x_1]$ is $C^1$-continuous. For the 3D point $P$ of $f\wedge g$ corresponding to a VT point, if we require the approximating space curve is $C^1$-continuous at $P$, then the whole approximating space curve is also $C^1$-continuous.
$\Box$

 When re-parameterizing the approximation space regular curve segments into rational ones, we need to know the tangent directions of the endpoints of space regular curve segments. For the endpoints corresponding to non-VT points, we can directly get it from the tangent directions of the plane curves. For the endpoints corresponding to VT points, we can get the tangent directions as follows. At first, we assume that $(x,p(x)),(x,q(x)),x\in [x_0,x_1]$ are parametric plane regular curve segments of exact algebraic regular curve segments $(x,\tilde{y}_1(x)),(x,\tilde{y}_2(x)),x\in [x_0,x_1]$ and $x_0$ corresponds to a VT point. The exact tangent direction of the algebraic space regular curve segment at $x_0$ is $(1,\tilde{y}_1'(x_0),\frac{\tilde{y}_2'(x_0)-\tilde{y}_1'(x_0)}{s}$ from the parametric representation. Note that $(1,\infty)$ corresponds to $(0,1)$ for plane regular curve segments. So for the approximation tangent direction at $x_0$: $(1,\frac{\partial p(x_0)}{\partial x},(\frac{\partial q(x_0)}{\partial x}-\frac{\partial p(x_0)}{\partial x})/s)$, if $\frac{\partial p(x_0)}{\partial x}$ is larger than (or less than) some given value, for example, 100 (or -100), we can reset the tangent direction as $(0,1,(\frac{\partial q(x_0)}{\partial x}-\frac{\partial p(x_0)}{\partial x})/(s\,\frac{\partial p(x_0)}{\partial x}))$. Moreover, if $(\frac{\partial q(x_0)}{\partial x}-\frac{\partial p(x_0)}{\partial x})/(s\,\frac{\partial p(x_0)}{\partial x})$ is larger than (or less than) some given value, we can set the tangent direction as $(0,0,\pm1)$. So the tangent directions at $x_0$ is as $(0,1,p),p\neq0$ or $(0,0,\pm 1)$.

\noindent{\bf Reparametrization of space curve.} If the tangent direction at $x_0$ is $(0,1,p)$, we can re-parameterize the space curve segment with the form
{\tiny
\begin{equation}\label{re-parameterize}
\mathbf{P}(t)=(\frac{a_1\,t^2+b_1\,t+c_1}{d_1\,t+1},t,\frac{a_2\,t^2+b_2\,t+c_2}{d_2\,t+1}+\frac{c_3}{d_3\,t+1}),\,t\in
[0,1],
\end{equation}
}
such that it is $G^1$-continuous with other regular curve segments at the endpoints. Assume that the two endpoints are
$(x_i,y_i,z_i), i=0,1$ and the given tangent directions at two endpoints are
$(x_i',y_i',z_i'), i=0,1 $. Thus $x_0'=0$. Here for simplicity, we assume that $y_0=0,y_1=1$ since we can set $t=\frac{y-y_0}{y_1-y_0}$.
Bisecting the regular curve segment ensures that $y_1'\neq0$ since the regular curve segment is monotonous. %And the value of $(x_1',y_1',z_1')$ can be derived from the formula $(1,\tilde{y}_1'(x_1),\frac{\tilde{y}_2'(x_1)-\tilde{y}_1'(x_1)}{s})$ for the $x$-coordinate $x_1$ of the endpoint. But for the value of $(0,y_0',z_0')$, we need to computing it by approximating from the tangent directions of the points on two plane projection curves. It is not difficult to find that the following formula holds.
%  $$\frac{z_0'}{y_0'}=\lim_{x\rightarrow x_0^+}\frac{\tilde{y}_2'(x)-\tilde{y}_1'(x)}{s\, \tilde{y}_1'(x)}.$$
%So we can use a rational number $\bar{x_0}>x_0$ but close to $x_0$ to approximate $\frac{z_0'}{y_0'}$. For example, let $\bar{x_0}=b$ such that $[a,b]$ is the isolating interval $x_0$.
%If $\frac{z_0'}{y_0'}=\pm\infty$,  $(0,y_0',z_0')=\lambda\,(0,0,\pm1)$.

We require that the parametric space curve
satisfying $G^0$ and $G^1$ conditions at the two endpoints. So we
have eight valid equations from the following equations.
{\tiny \begin{eqnarray*}
&&\mathbf{P}(t)|_{t=0}=(x_0,y_0,z_0),\mathbf{P}(t)|_{t=1}=(x_1,y_1,z_1),\\
&&\frac{\partial \mathbf{P}(t)}{\partial t}|_{t=0}=(0,1,p), \frac{\partial \mathbf{P}(t)}{\partial t}|_{t=1}=\frac{1}{y_i'}(x_1',y_1',z_1').
\end{eqnarray*}
}
Solving them, we have one solution as below.
{\tiny
\begin{eqnarray*}
a_{{1}}&=&{\frac {{x_{{0}}}^{2}-2\,x_{{0}}x_{{1}}+{x_{{1}}}^{2}}{-x_{{1
}}+x_1'+x_{{0}}}},\\
a_{{2}}&=&\frac{1}{d_3}( z_1'd_{{3}}+z_1'd_{{2
}}d_{{3}}-z_{{1}}d_{{3}}+d_{{3}}z_{{0}}+z_0'+z_1'd_{{2}}-z_{
{1}}d_{{2}}+d_{{2}}z_{{0}}+z_1'\\
&&-2\,z_{{1}}+2\,z_{{0}} ),\\
b_{{1}}&=&-{\frac {x_{{0}} \left( 2\,x_{{0}}-2\,x_{{1}}+x_1'
 \right) }{-x_{{1}}+x_1'+x_{{0}}}},\\
b_{{2}}&=&-\frac{1}{d_3^2}(-z_{{1}}d_{{2
}}+d_{{2}}z_{{0}}+z_1'd_{{2}}-2\,z_{{1}}d_{{2}}d_{{3}}+2\,d_{{2}}
z_{{0}}d_{{3}}+z_1'd_{{2}}{d_{{3}}}^{2}\\
&&+2\,z_1'd_{{2}}d_{{3}
}-z_{{1}}d_{{2}}{d_{{3}}}^{2}+4\,d_{{3}}z_{{0}}+2\,z_{{0}}-2\,z_{{1}}
+ z_0'+z_1'{d_{{3}}}^{2}+2\,z_1'd_{{3}}\\
&&-4\,z_{{1}}d_{{3}}
+2\,d_{{3}}z_0'+z_1'-2\,z_{{1}}{d_{{3}}}^{2}
+2\,{d_{{3}}}^{2
}z_{{0}}),\\
c_{{1}}&=&x_{{0}},\\
c_{{2}}&=&-\frac{1}{{d_3}^{2}
 ( -d_{{3}}+d_{{2}} )} ( -2\,z_{{1}}
d_{{2}}d_{{3}}+z_1'd_{{2}}{d_{{3}}}^{2}+2\,z_1'd_{{2}}d_{{3}
}+z_1'
+z_0'-2\,z_{{1}}\\
&&+2\,z_{{0}}-z_{{1}}d_{{2}}-4\,z_{{1}}d_{{3}}
+z_1'd_{{2}}+z_1'{d_{{3}}}^{2}+2\,z_1'd_{{3}}-z_{{1}}d_{{2}}{d_{{3}}}^{2}+2\,d_{{2}}z_{{0}}d_{{3}}\\
&&
+d_{{2}}z_{{0}}+4\,d_{{3}}z_{{0}}+2\,{d_{{3}}}^{2}z_{{0}}
+2\,d_{{3}}z_0'+{d_3}^{2}z_0'
-2\,z_{{1}}{d_{{3}}}^{2}+{d_{{3}}}^{3}z_{{0}}),\\
c_3&=&\frac{1}{{d_3}^{2}
 ( -d_3+d_2 )} (-2\,z_{{1}}d_{{2}}
d_{{3}}+z_1'd_{{2}}{d_{{3}}}^{2}+2\,z_1'd_{{2}}d_{{3}}+z_1'+z_0'-2\,z_{{1}}\\
&&+2\,z_{{0}}-z_{{1}}d_{{2}}-4\,z_{{1}}d_{{3}}
+z_1'd_{{2}}+z_1'{d_{{3}}}^{2}+2\,z_1'd_{{3}}-z_{{1}}d_{
{2}}{d_{{3}}}^{2}+d_{{2}}{d_{{3}}}^{2}z_{{0}}\\
&&+2\,d_{{2}}z_{{0}}d_{{3}}
+d_{{2}}z_{{0}}+4\,d_{{3}}z_{{0}}+2\,{d_{{3}}}^{2}z_{{0}}+2\,d_{{3}}z_0'+{d_{{3}}}^{2}z_0'-2\,z_{{1}}{d_3}^{2}),\\
 d_{{1}}&=&-{\frac {2\,x_{{0}}-2\,x_{
{1}}+x_1'}{-x_{{1}}+x_1'+x_{{0}}}},
\end{eqnarray*}
}
where $d_2,d_3$ are free variables. At first, we require that ${\it x_1'}-{\it x_1}+{\it x_0}\neq 0$,  $d_3(d_2-d_3)\neq 0$ since they are denominators. Second, we require that $d_i\,t+1=0 (i=1,2,3)$ in $t$ has no root in $[0,1]$, that is, $d_i>-1$. For $i=1$, we have equal conditions:
\begin{equation}\label{eq-cond}(x_0-x_1)(x_0-x_1+x_1')<0.\end{equation}
Since the given planar regular curve segment is monotonous (w.r.t. both $x$ and $y$),  the first condition and (\ref{eq-cond}) hold directly. We can choose proper $d_2,d_3$ such that conditions hold.

%To ensure that the approximation is fine, we require
%that for any $x\in[x_0,x_1]$, the equation $x(t)=x$ in $t$ has only one $t\in[0,1]$. Modifying the equation, we have an equation in $t$ with degree 2, say $p(t)=0$. If $p(0)p(1)<0$, the required condition is satisfied, where
%$$L=p(0)p(1)=-(-x_1+x_1'+x_0)\,(-x_0+x)\,(x_0-x_1)\,(-x_1+x).$$
%Since $(x-x_0)\,(x-x_1)<0$ and $x_1-x_0>0$,  $L<0$ can be simplified as
%$$(\frac{x_1'}{x_1-x_0}-1)(\frac{x_0'}{x_1-x_0}-1)e_1^2+(\frac{x_1'}{x_1-x_0}-1)e_1<0.$$
%From the way we construct the plane regular curve segments of $h=0$ and $\bar{h}=0$, we know that they are monotonous w.r.t. $x$. So do the space regular curve segments. Thus $(\frac{x_1'}{x_1-x_0}-1)(\frac{x_0'}{x_1-x_0}-1)<0$. From the property of quadric equation, we can always find $e_1$ satisfying the required conditions. In order to find well-defined rational functions for the parametric space curve, we also require that $t^2+d_i\,t+e_i=0, i=1,2,3$ have no real roots in $[0,1]$. So in the end, we can obtain a $C^1$-continuous quadric rational approximating curve for $f\wedge g$.

For the case of tangent direction is $(0,0,1)$, we can set the parametric regular curve segment as
{\tiny
\begin{eqnarray*}
\mathbf{P}(t)=(\frac{a_1\,t^2+b_1\,t+c_1}{d_1\,t+1},\frac{a_2\,t^2+b_2\,t+c_2}{d_2\,t+1}+\frac{c_3}{d_3\,t+1},t),\,t\in
[0,1],
\end{eqnarray*}
}
and solve a similar equation system to get the parametric regular curve segments.

The left problem is to control the precision. Let $\epsilon$ be the required precision for the whole approximation parametric curve. If the non-rational parametric curve $\mathcal{S}_1: (x,p(x),q(x)), x\in [x_0,x_1]$ approximates the regular curve segment of algebraic space curve $S$ with precision $\epsilon/2$, and the new rational parametric curve $\mathcal{S}_2: (x(t),y(t),z(t))$ approximate $\mathcal{S}_1: (x,p(x),q(x))$ with precision $\epsilon/2$, then $\mathcal{S}_2$ approximate $S$ with precision $\epsilon$.

We need to control the approximation precision of $\mathcal{S}_2$ to $\mathcal{S}_1$. In \cite{shen}, the authors consider the approximation of 3-D parametric curve with rational B$\acute{e}$zier curves. For our problem, we need rational curve. For any fixed $x^0\in[x_0,x_1]$, we can derive a univariate polynomial equation in $t$ of degree 2 by $p(x^0)=y(t)$. Solving it, we have two real solutions (the solutions do exist). Choose the one such that $x(t)$ close to $x^0$, say $t^0$. Denote the distance between $(x^0,p(x^0),q(x^0))$ and $(x(t^0),y(t^0),z(t^0))$ as $D(x_0)$. From the definition (\ref{eq-herr}), we can find that $\max_{x^0\in[x_0,x_1]}D(x^0)\ge e(\mathcal{S}_1,\mathcal{S}_2)$ is an upper bound of the Hausdorff distance of $\mathcal{S}_1$ and $\mathcal{S}_2$. We can choose some sample points to estimate the error between $\mathcal{S}_1$ and $\mathcal{S}_2$.

Thus, in the end, we get a $G^1$-continuous piecewise rational approximation space curve under a given precision.

When we approximate a regular curve segment containing a VT point in practice, we usually select a short distance for it since the error control is much easier.

\section{Algorithm and examples}
In this section, we will give the main algorithm to approximate algebraic space curves and use some non-trivial examples to illustrate the effectivity of our algorithm.
\begin{algo} The inputs are $f, g\in \Q[x,y,z]$ such that $\gcd(f,g)=1$ and satisfying the two assumptions, a bounding box
$\B=[X_1,X_2]\times[Y_1,Y_2]\times[Z_1,Z_2]$ and an error bound $\epsilon >0$. The outputs are piecewise rational parametric regular curve segments $\C_i:=\{(x,y_i(x),z_i(x)) (\hbox{ or } (x_i(y),y,z_i(y)),a_i\le x (\hbox{ or } y ) \le b_i, (i=1,...,N)\}$, which give a $G^1$-continuous approximation to $f\wedge g$ in $\B$ with precision $\epsilon$.
\end{algo}
\begin{enumerate}
\item Topology determination and regular curve segmentation of the plane curve defined by $\mathcal{C}_1: \pi_z(f\wedge g)$.
\item Compute a rational number $s$ as mentioned in Theorem \ref{thm-correspondence}.
\item Let $\varphi_s: (x,y,z)\rightarrow (x,y+s\,z,z)$. Topology determination  and regular curve segmentation of the plane curve defined by $\mathcal{C}_2: \pi_z(\varphi_s(f)\wedge \varphi_s(g))$.
\item Find out the correspondence between the regular curve segments of $\mathcal{C}_1$ and $\mathcal{C}_2$.
\item Approximate the regular curve segments without VT point of $\mathcal{C}_1$ and $\mathcal{C}_2$ with $\epsilon_0<\frac{s}{\sqrt{s^2+4}}\epsilon$ and the ones with VT point with precision $\epsilon_0<\frac{s}{2\sqrt{s^2+4}}\epsilon$.
\item Recover the space approximation regular curve segments of $f\wedge g$ with formula (\ref{eq-app}).
\item Re-parameterize the non-rational approximation curves to rational approximation curves under the error control if there exist.
\item Output the piecewise approximation regular curve segments. %preserving the topology and satisfying the given precision $\epsilon$.
\end{enumerate}
%\noindent{\bf Proof.} The correctness of the algorithm is guaranteed by Theorem \ref{thm-correspondence}, Lemma \ref{lem-continuous} and Theorem \ref{lem-err} for topology, $G^1$-continuousness and error control, respectively. The termination of the algorithm is guaranteed by the error convergence to zero by subdividing the regular curve segments recursively. $\Box$

\begin{figure}[hb]\label{fig-example1}
\begin{minipage}{0.234\textwidth}
\centering
\includegraphics[height=25mm]{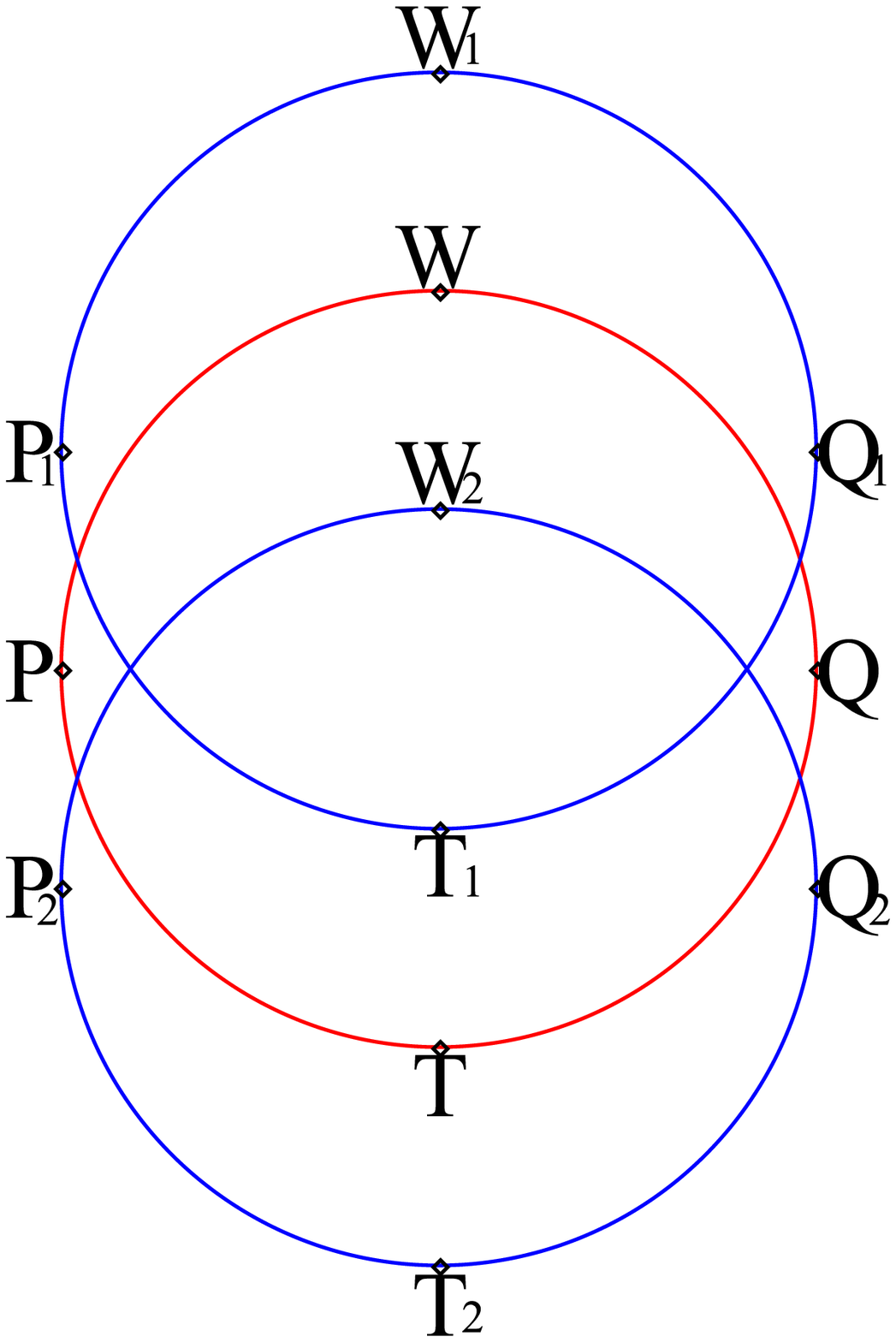}
\end{minipage}
\begin{minipage}{0.234\textwidth}
\centering
\includegraphics[height=25mm]{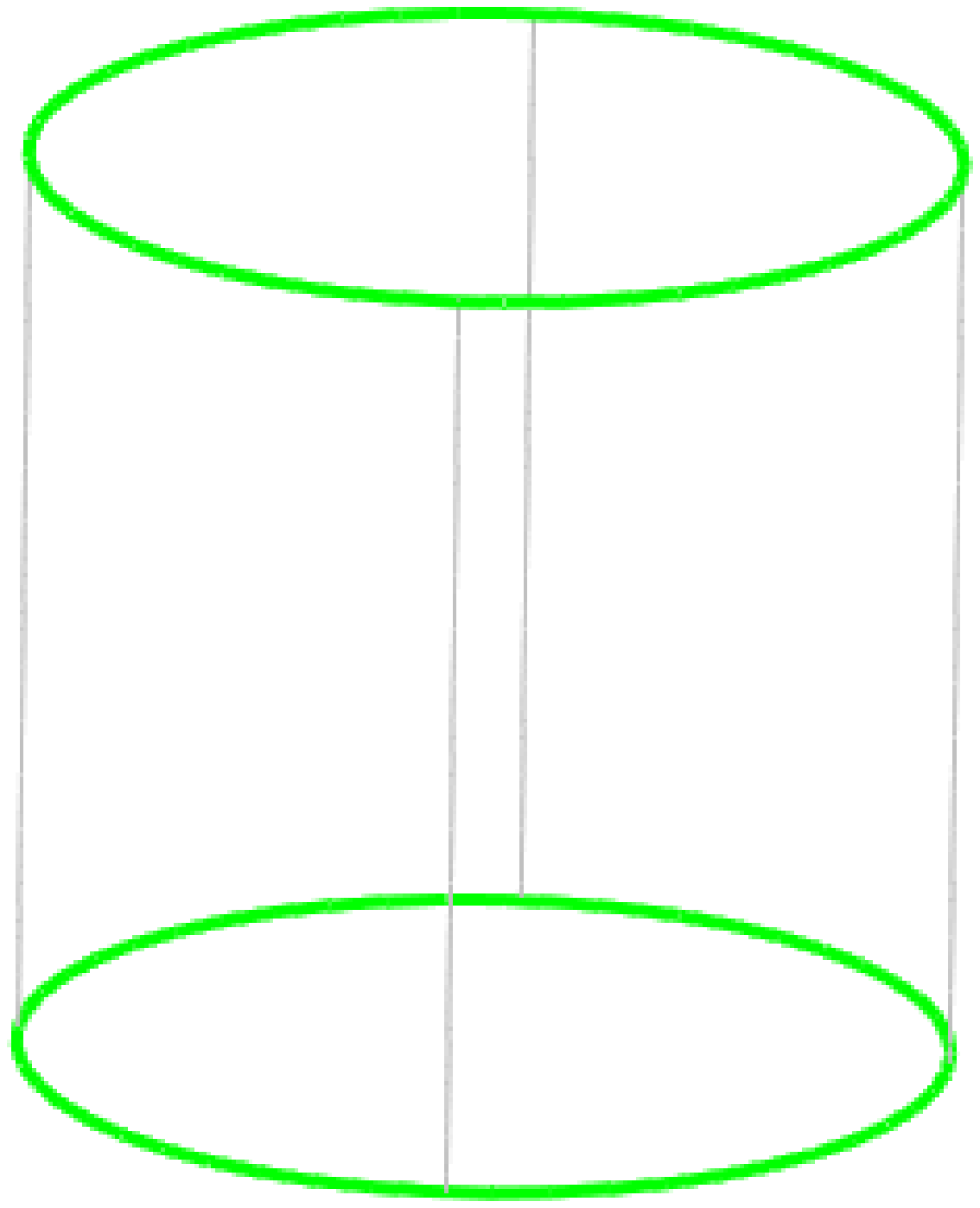}
\end{minipage}
\caption{Projection curves and approximation curve in Example \ref{ex-1}.}
\end{figure}

We will show several examples to illustrate our algorithm.

\begin{exmp}\label{ex-1}
Consider the algebraic space curve defined by the system $\{f,g\}=\{x^2+y^2+z^2-4,(z-1)\,(x^2+y^2-3\,z^2)\}$. In fact, they are two plane circles with $z=\pm 1$ as shown in Figure 7 (green ones).
%This two circles overlap when projected to $xy$-plane and the circle with $z=1$ is not regular.
The space curve is not irreducible, not regular, and not in a generic position.
We will approximate it with rational curves under precision $10^{-2}$. Following Algorithm 7, we have
\begin{enumerate}
\item Compute the resultant of $f,g$ w.r.t. $z$, we have $\Cc_1: h=x^2+y^2-3=0$, as the red circle in Figure 7. We split $\Cc_1$ into eight regular curve segments with $x$-coordinates $[-1.732050808,-1.0,0,1.0,1.732050808]$:
%{\tiny $[[(-1.732050808,0),(-1,-1.414213562)]$, $[(-1,-1.414213562),(0,-1.732050808)]$, $[(0,-1.732050808),(1,-1.414213562)]$, $[(1,-1.414213562)$, $(1.732050808,0)]$,
%    $[(-1.732050808,0),(-1,1.414213562)]$, $[(-1,1.414213562),(0,1.732050808)]$, $[(0,1.732050808)$, $(1,1.414213562)]$, $[(1,1.414213562),(1.732050808,0)]]$.}
    And the tangent directions of the points all are $(1,-\frac{\partial h}{\partial x}/\frac{\partial h}{\partial y})$ evaluated at these points.  Note that $x=0$ corresponds $x$-critical points of $\Cc_1$ and $(\pm1.732050808,0)$ correspond to VT points.

\item Since $\pi_y(h)=x^2-3$, we can obtain\\ {\tiny $\alpha_0=-2,\alpha_1=-1.732050808,\alpha_2=0,\alpha_3=1.732050808$, $\alpha_4=2$.} And we can get $r=3.464101616$. Following Theorem \ref{thm-correspondence}, we have $R=1.0$ when we choose $g$ to compute $R$. We can select $s=1<\frac{r}{2R}=1.732050808$.

\item Compute the resultant of $\varphi(f)=f(x,y+z,z), \varphi(g)=g(x,y+z,z)$ w.r.t. $z$, we have $\Cc_2: \bar{h}=(x^2+y^2-2+2\,y)\,(-2+x^2-2\,y+y^2)=0$, as two blue circles in Figure 7. Since $\pi_y(\bar{h})=(x^2-3)\,(x^2-2)=0$, we split $\Cc_2$ into 16 regular curve segments at\\ {\tiny $x=\{-1.732050808,-1.414213562,0$, $1.414213562,1.732050808\}$.} And the tangent directions at the endpoints of these regular curve segments are $(1,\frac{\partial \bar{h}}{\partial x}/\frac{\partial \bar{h}}{\partial y})$ evaluated at the points close to these points. We can get the approximating tangent directions. And we can find that $(\pm1.732050808,\pm 1)$ are VT points since the absolute values of $\frac{\partial \bar{h}}{\partial x}/\frac{\partial \bar{h}}{\partial y}$ evaluated at $(\pm1.73204,\pm1.006118660),(\pm1.73204, \pm.9938813400)$ are larger than 200.
    %Following the method in \cite{gaoli}, we can compute the tangent directions are $(1,\pm 1.414213562), (1,\pm 1.414213562)$ respectively.

\item As shown in Figure 7, the critical points of $\Cc_1$ are $P,Q$. Choose a vertical line which intersect $\Cc_1$ at $W,T$. $K (K=P,Q,W,T)$ are points on $\Cc_1$ and $K_1,K_2$ are corresponding points of $K$ on $\Cc_2$. Consider {\tiny $W(0, 1.732050808), W_1(0, 2.732050808), W_2(0, 0.732050808)$} for example.  We can find that $W_1,W_2$ are on the line $x=0$ in a neighborhood with radius $1.732050808$ centered at $W$. So we can conclude that $W_1, W_2$ correspond to $W$ with local generic position method. The correspondence of other points are similar.

\item Approximate $\Cc_1,\Cc_2$ respectively. In order to derive the required precision $10^{-2}$, we we use precision $\epsilon_1=0.0044<\frac{1}{\sqrt{1^2+4}}10^{-2}$ for the regular curve segments of $\Cc_1, \Cc_2$ without VT point(s), and we use precision $\epsilon_2=0.0022<\frac{1}{2\sqrt{1^2+4}}10^{-2}$ for the regular curve segments with VT point. Consider
    %two regular curve segments in $\Cc_1,\Cc_2$ respectively for example. For the
    a regular curve segments on $\Cc_1$, $(-1.732050808,0), (-1.60,0.6633249580)$ are the endpoints for the one, denoted as $C_1$. And it has a VT point. $(-1.60,0.6633249580), (-1.40, 1.019803903)$ are endpoints for the other, denoted as $C_2$. And it has no VT point.
The approximation of $C_1$ is {\tiny $(x,0.9999999059\,\sqrt{-x^2+0.000000464\,x+3.000000806}), x\in
[-1.732050808,-1.60]$} and the error is very small. The approximation for $C_2$
is {\tiny $(x, 0.6106757885\,x+ 2.310809554 -0.1270414345/(0.5070598449\,x+ 1.0),x\in [-1.60,-1.40]$} and the error is
$0.0004<\epsilon_1$. For the regular curve segments on $\Cc_2$ with endpoints:
{\small $[(-1.732050808, 1.0)$,  $(-1.60, 1.663324958)]$},
denoted as $C_3$ and it has a VT point. $(-1.60$, $1.663324958)$,
$(-1.414213562, 2.0)$ are endpoints for the other, denoted as $C_4$, without
VT point. Similarly as $C_1$, the approximation for $C_3$ is
{\tiny $(x,1.0+0.9999999054\,\sqrt{-x^2+0.000000466\,x+3.000000809},x\in [-1.732050808,-1.60].$} The approximation for
$C_4$ is {\tiny $(x,0.6301674345\,x+3.324059999-0.1265242054/(0.5086364591\,x+ 1.0)),x\in [-1.60,-1.414213562]$} and the error is $
0.0002<\epsilon_1$. We can find that parts of $C_1, C_2$ and
$C_3,C_4$ are correspondent.

\item Recover the approximation space curves of $f\wedge g$ by the formula $z=\frac{y_2(x)-y_1(x)}{s}$. The space regular curve segment corresponding to $C_1$ and $C_3$, we have its approximation parametric space regular curve segment for  $x\in [-1.732050808,-1.60]$: {\tiny \begin{eqnarray*} &&(x,0.9999999059\,\sqrt{-x^2+0.000000464\,x+3.000000806},\\
    &&1.0+0.9999999054\,\sqrt{-x^2+0.000000466\,x+3.000000809}\\
    &&-0.9999999059\,\sqrt{-x^2+0.000000464\,x+3.000000806}).\end{eqnarray*}}
The approximation space curve is not rational, denoted as $S_1$. The approximation
corresponding to $C_2, C_4$ for $x\in [-1.60,-1.414213562]$, denoted as $S_2$, is\\
%   {\tiny \begin{eqnarray*} &&(x, 0.6106757885\,x+ 2.310809554 -\frac{0.1270414345}{0.5070598449\,x+ 1.0},\\
%    &&0.0194916460\,x+1.013250445-\frac{0.1265242054}{0.5086364591\,x+1.0}\\
%    &&+\frac{0.1270414345}{0.5070598449\,x+1.0}).\end{eqnarray*}}
   {\tiny $(x, 0.6106757885\,x+ 2.310809554 -\frac{0.1270414345}{0.5070598449\,x+ 1.0},$\\
    $0.0194916460\,x+1.013250445-\frac{0.1265242054}{0.5086364591\,x+1.0}+\frac{0.1270414345}{0.5070598449\,x+1.0}).$}

%It is rational, .
% We denote the two approximation space regular curve segments as $S_1,S_2$ respectively.

\item We will re-parameterize $S_1$ into rational one.
%Since a rational representation as $(x,y(x),z(x))$ for $S_1$ is not
%obtainable,
%We assume the rational approximation regular curve segment for
%$S_1$ is as Formula (\ref{re-parameterize}).
%$(x,y,z)=(\frac{a_{1}t^2+b_{1}t+c_{1}}{d_{1}t+1},t,1), t\in[0,\frac{\sqrt{11}}{5}]$.
At first, we can find that the $y$ coordinate of $S_1$ changes from $0$ to $0.6633249580$. Its two endpoints are $P_0(-1.732050808,0,1.0), P_1(-1.60,0.6633249580,1.0)$. The tangent direction of $S_2$ at $P_2$ is $(1,2.412090757,0.0)$.
By approximating the tangent direction of $S_1$ at $P_1$, we have $(1,283.0783218,0.0)$. And there is another regular curve segment which shares a same tangent direction with $S_1$ at $P_1$. Taking their average value, we can set the tangent direction of $S_1$ at $P_1$ as $(0,1,0)$. Using Formula
\bref{re-parameterize}, we can easily obtain the rational approximation regular curve segment for $S_1$ is\\
%{\tiny \begin{eqnarray*}(-2.412090984\,y-22.71853197+\frac{20.98648116}{-0.1149354656\,y+1},y,1.0)\end{eqnarray*}}
{\tiny $(-2.412090984\,y-22.71853197+\frac{20.98648116}{-0.1149354656\,y+1},y,1.0).$}\\
The error in $x$-direction is $0.0020563160<\epsilon_2$, (We take 19
sample points besides endpoints to compute the error.). So the
approximation rational curve satisfies the error requirement.

\item Output the piecewise approximation curve. %The approximation space curve is as shown in Figure 7 (green ones).
\end{enumerate}
\end{exmp}

%We compute two more examples to illustrate our algorithm and they work well.
\begin{exmp} \label{ex-2} Approximate the algebraic space curve defined by $f=g=0$, where
$f={x}^{2}+{y}^{2}+{z}^{2}-4, g=\left( {x}^{2}+{y}^{2}+2\,y-{z}^{2}
 \right)  \left( z-x-4\,y \right)$. It is a space curve with singular point. The approximation space curve is as the left part of Figure 8 %\ref{fig-singular}
 and the error is 0.013. The color differs the different approximating space regular curve segments.
\end{exmp}

    \begin{figure}[hb]\label{fig-singular}
\begin{minipage}{0.234\textwidth}
\centering
\includegraphics[height=32mm]{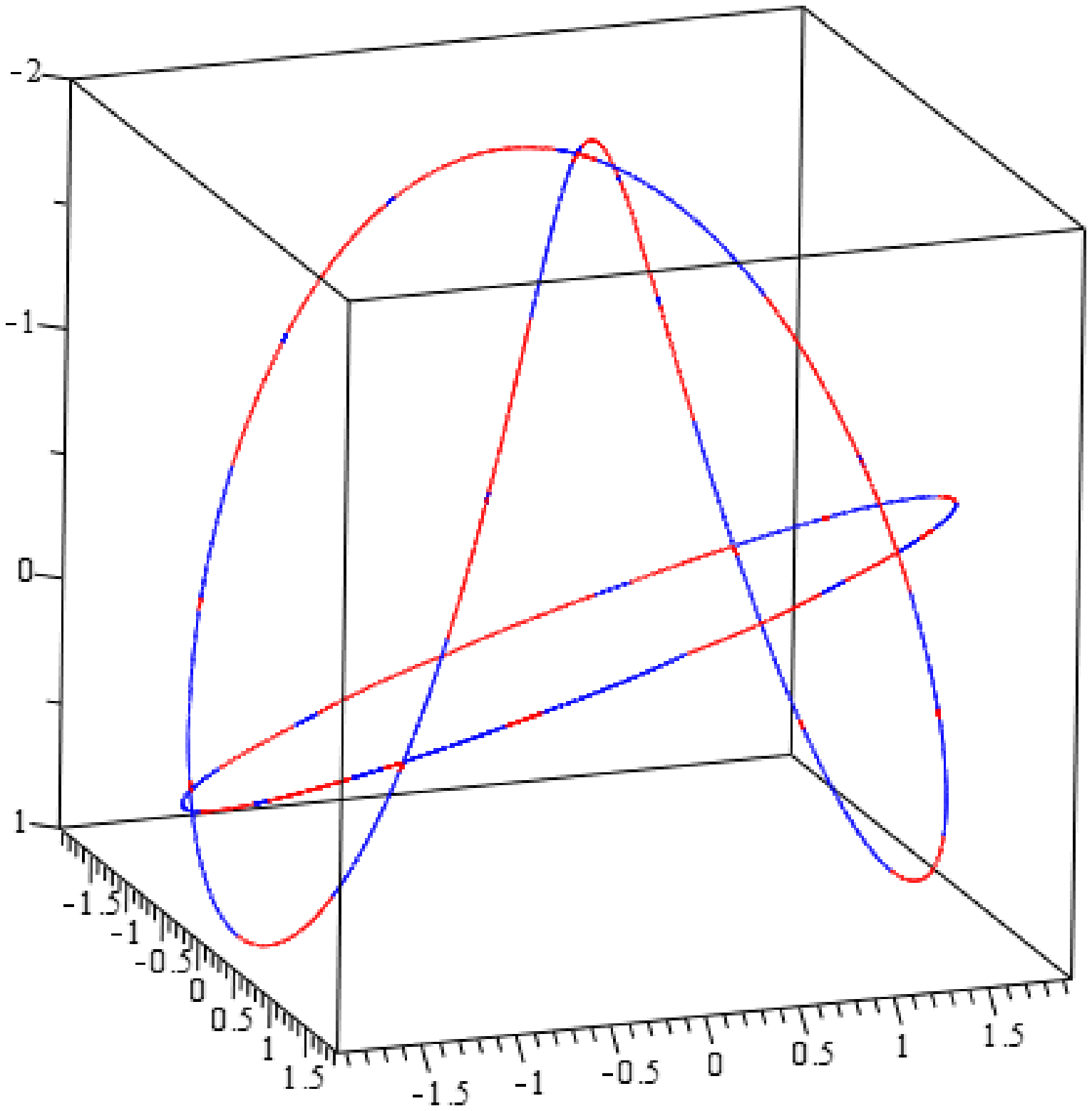}
\end{minipage}
\begin{minipage}{0.234\textwidth}
\centering
\includegraphics[height=24mm]{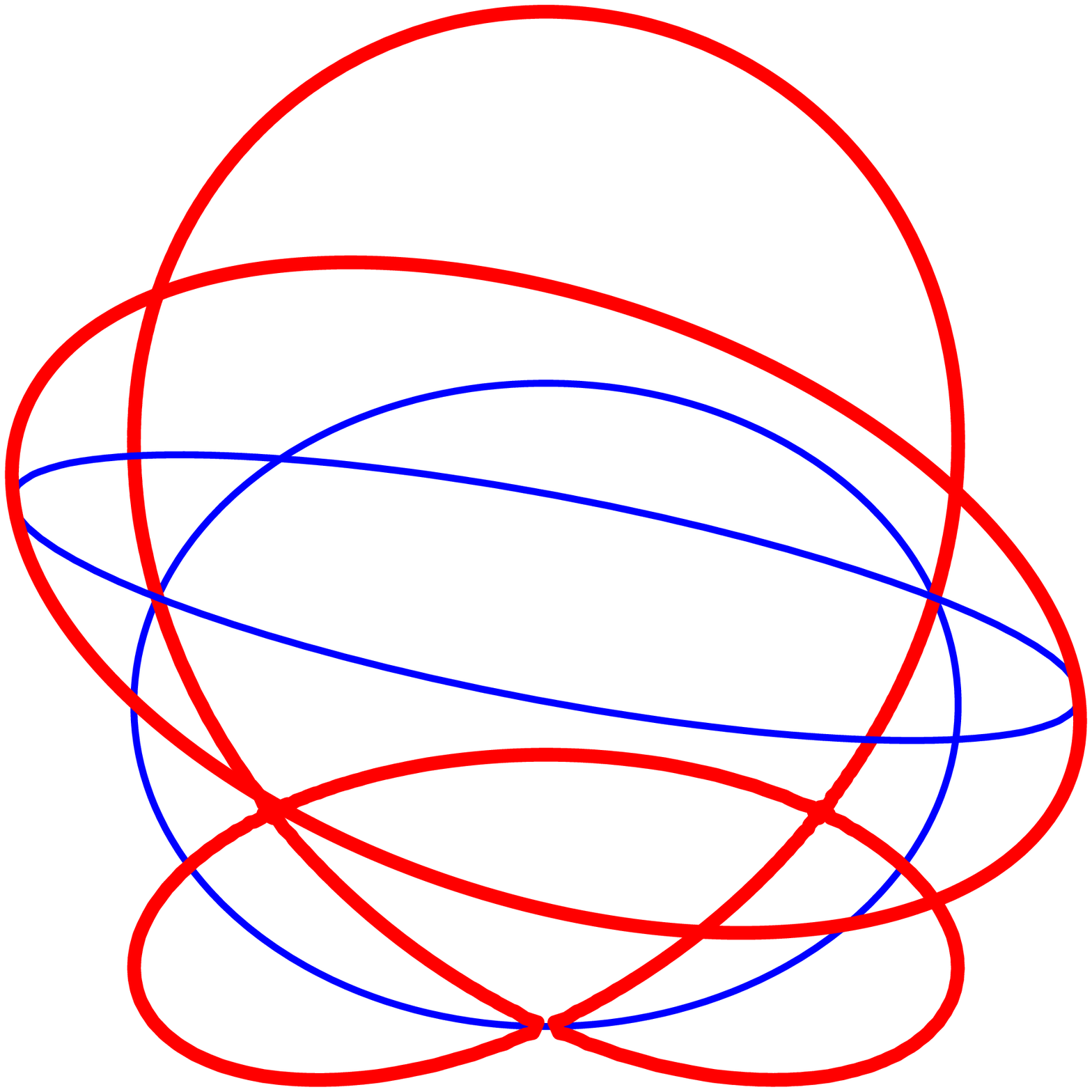}
\end{minipage}
\caption{Appriximation curve and projection curves in Example \ref{ex-2}.}
    \end{figure}

\begin{exmp}\label{ex-3} In this example, we will approximate the algebraic space curve defined by $f=g=0$ inside $[-2,2]\times [-2,2]\times[-2,2]$ with error $\epsilon=0.014$, where
$f=95-94\,{x}^{3}+64\,{x}^{2}y+28\,{x}^{2}z-61\,{x}^{2}+69\,x{y}^{2}-53
\,xyz-59\,xy+28\,x{z}^{2}-15\,xz-83\,x-3\,{y}^{3}+59\,{y}^{2}z+49\,{y}
^{2}+4\,y{z}^{2}+11\,yz+5\,y-81\,{z}^{3}-8\,{z}^{2}-9\,z, g=49+7\,{x}^{3}
-46\,{x}^{2}y+87\,{x}^{2}z+94\,{x}^{2}+73\,x{y}^{2}+93\,xyz-3\,xy-27\,
x{z}^{2}+56\,xz+70\,x+72\,{y}^{3}-37\,{y}^{2}z-20\,{y}^{2}+79\,y{z}^{2
}-78\,yz-3\,y+94\,{z}^{3}+30\,{z}^{2}+47\,z
$. The approximation space curve is as Figure 9.%\ref{fig-regular}.
\end{exmp}

    \begin{figure}[hb]\label{fig-regular}
\begin{minipage}{0.234\textwidth}
\centering
\includegraphics[height=32mm]{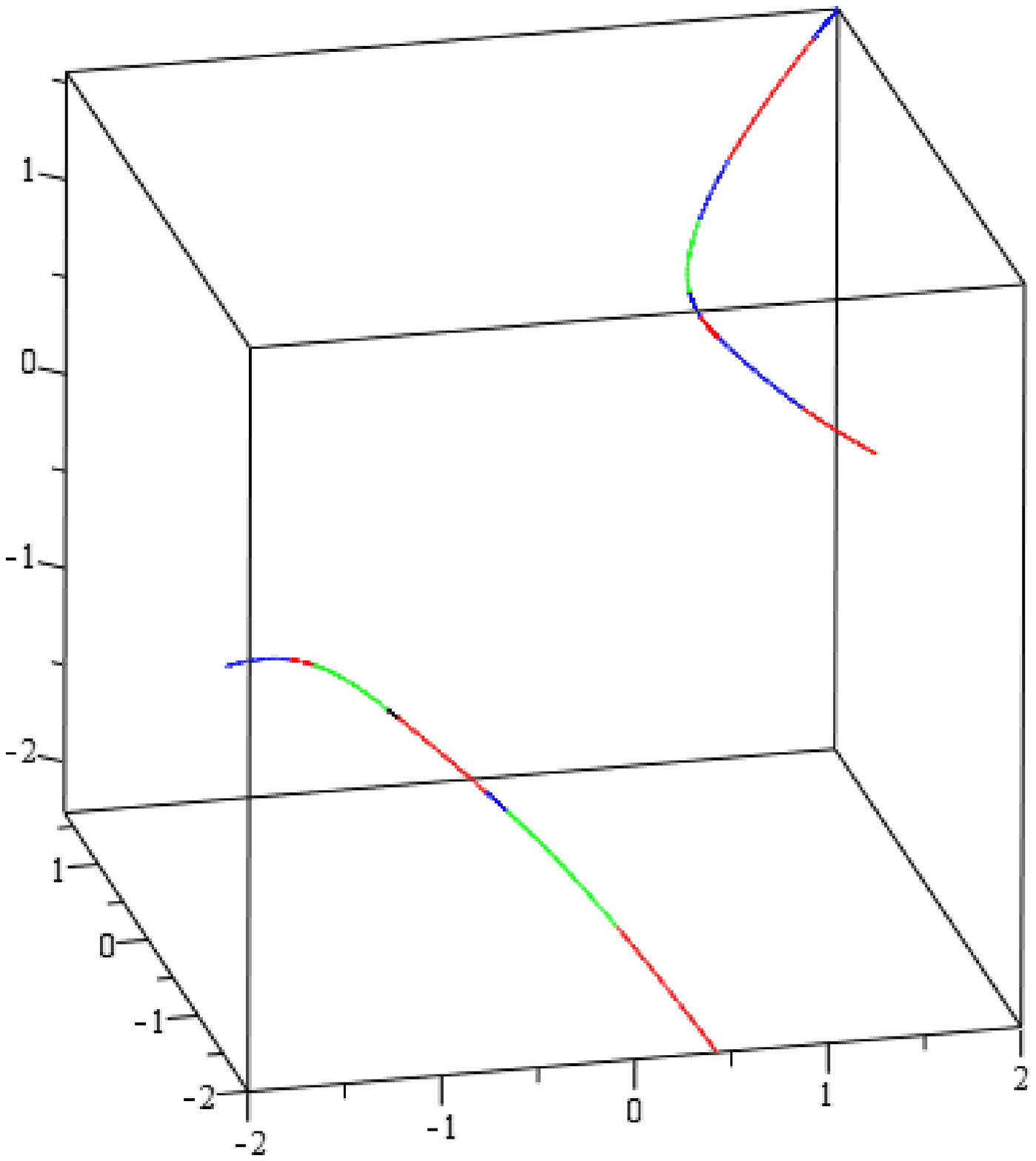}
\end{minipage}
\begin{minipage}{0.234\textwidth}
\centering
\includegraphics[height=25mm]{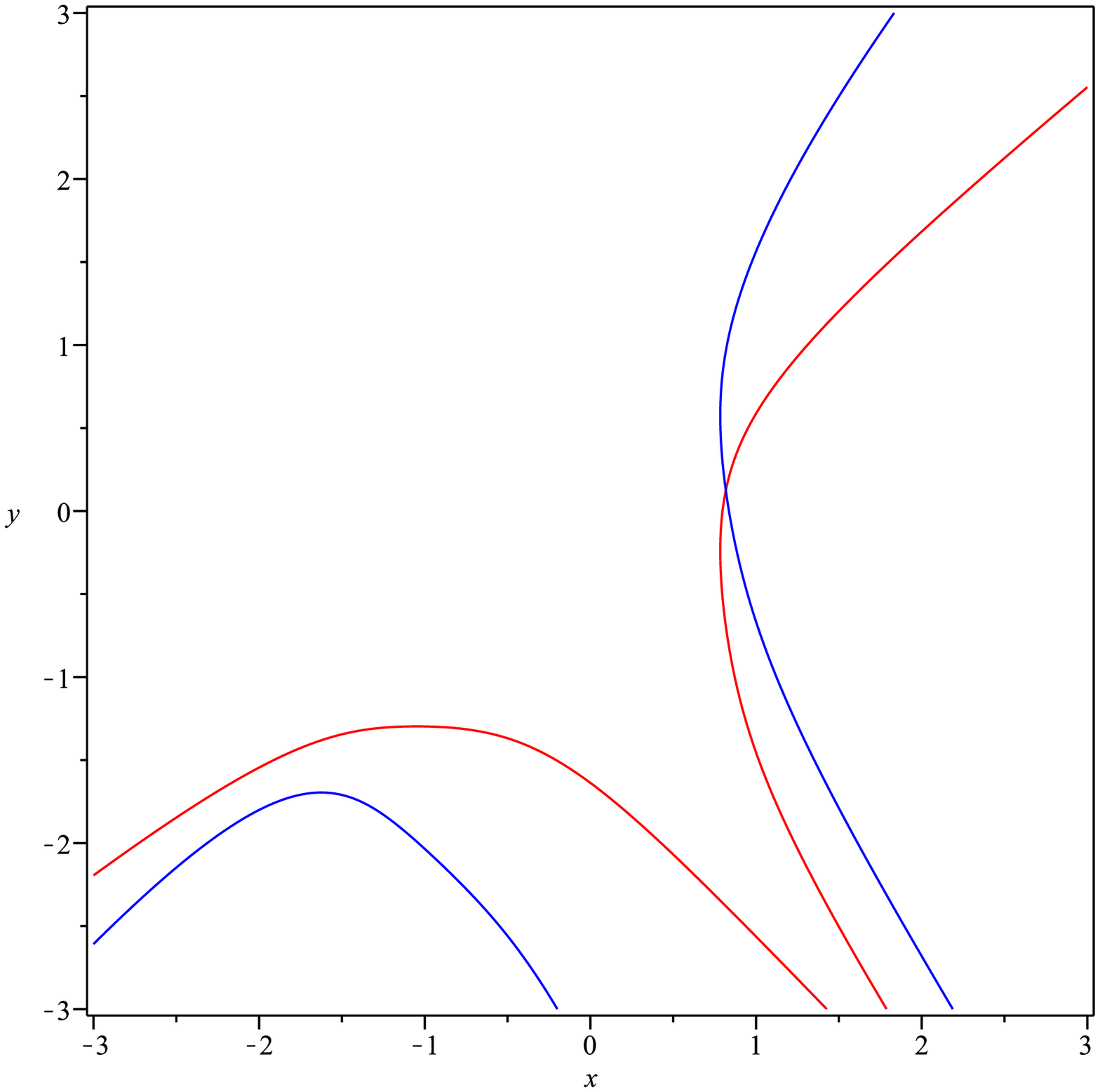}
\end{minipage}
\caption{Appriximation curve and projection curves in Example \ref{ex-3}.}
    \end{figure}

\section{Conclusion}
We introduce a local generic position method to compute the topology as well as the piecewise approximation curves of algebraic space curves. Especially, we present an algorithm to approximate algebraic space curve by piecewise rational curves with correct topology and under any given precision. The method is effective.

\section*{Acknowledgement} The work is partially supported by NKBRPC
(2011CB302400), NSFC Grants (60821002, 11001258, 91118001), and China-France
cooperation project EXACTA (60911130369).

\end{document}